\documentclass[final,3p,times]{elsarticle}
\usepackage{lineno,hyperref}
\usepackage{amssymb}
\usepackage{amsmath}
\usepackage{amsthm}
\usepackage{mathrsfs}
\usepackage{bm}
\usepackage{tikz}
\usepackage{pgfplots}
\usepackage[ruled,vlined]{algorithm2e}
\usepackage{subfig}
\usepackage{algorithmic,algorithm2e,float}
\usepackage{bbm}
\usepackage[T1]{fontenc}
\usepackage[utf8]{inputenc}
\usepackage[font=small,labelfont=bf]{caption}

\graphicspath{{./figs/}}
\usepackage{mathtools}

\newcommand{\myfrac}[2]{\displaystyle \frac{#1}{#2}}

\modulolinenumbers[100]
\journal{\textbf{Journal of Computational Physics, doi: https://doi.org/10.1016/j.jcp.2021.110683}}
\def\etal{{\it et al.~}}

\begin{document}

\begin{frontmatter}

\title{Parallel Physics-Informed Neural Networks via Domain Decomposition}


\author{Khemraj Shukla, Ameya D. Jagtap, George Em Karniadakis$^*$}
\cortext[mycorrespondingauthor]{Corresponding author Email:   george$\_$karniadakis@brown.edu, \\ 
}

\address{Division of Applied Mathematics, Brown University, 182 George Street, Providence, RI 02912, USA}

\begin{abstract}
We develop a distributed framework for the physics-informed neural networks (PINNs) based on two recent extensions, namely conservative PINNs (cPINNs) and extended PINNs (XPINNs), which employ domain decomposition in space and in time-space, respectively. 
This domain decomposition endows cPINNs and XPINNs with several advantages over the vanilla PINNs, such as parallelization capacity, large representation capacity, efficient hyperparameter tuning, and is particularly effective for multi-scale and multi-physics problems. 
Here, we present a parallel algorithm for cPINNs and XPINNs constructed with a hybrid programming model described by MPI $+$ X, where X $\in \{\text{CPUs},~\text{GPUs}\}$. The main advantage of cPINN and XPINN over the more classical data and model parallel approaches is the flexibility of optimizing all hyperparameters of each neural network separately in each subdomain.
We compare the performance of distributed cPINNs and XPINNs for various forward problems, using both weak and strong scalings.
Our results indicate that for space domain decomposition, cPINNs are more efficient in terms of communication cost but XPINNs provide greater flexibility as they can also handle time-domain decomposition for any differential equations, and can deal with any arbitrarily shaped complex subdomains. To this end, we also present an application of the parallel XPINN method for solving an inverse diffusion problem with variable conductivity on the United States map, using ten regions as subdomains. 
\end{abstract}


\end{frontmatter}

\linenumbers

\section{Introduction}
In recent years there has seen a surge of machine learning based techniques applied to many complex systems such as autonomous vehicles \cite{bojarski2016end}, speech recognition \cite{hinton2012deep, huang2014historical}, medical diagnosis \cite{litjens2017survey}, consumer preferences~\cite{khandani2010consumer}, etc. However, such techniques are less explored in the field of scientific computations. Physics-informed neural networks (PINNs) \cite{raissi2019physics} is a recently proposed deep learning method, which bridges the gap between machine learning based methods and scientific computations. Due to their simplicity, PINNs have led to progress in many areas of computational science, see for examples \cite{mao2020physics, shukla2020physics, sahli2020physics, yin2021non, waheed2020eikonal, shukla2021physics, VPINN, cai2021flow,cai2021artificial}. PINNs can efficiently tackle both forward problems, where the solution of governing physical law is inferred, as well as ill-posed inverse problems, where the unknown physics and/or free parameters in the governing equations are identified from the available multi-modal measurements. 

However, one of the major limitations of PINNs is the large computational cost associated with the training of the neural networks, especially for forward multi-scale problems. To reduce the computational cost, Jagtap \etal \cite{jagtap2020conservative} introduced a domain decomposition-based PINN for conservation laws, namely conservative PINN (cPINN), where continuity of the states as well as their fluxes across subdomain interfaces is used to stitch the subdomains together. In subsequent work, Jagtap \& Karniadakis \cite{jagtap2020extended} applied domain decomposition to general PDEs using the so-called extended PINN (XPINN). Unlike cPINN, which offers space decomposition, XPINN offers both space-time domain decomposition for any irregular, non-convex geometry thereby reducing the computational cost effectively. By exploiting the decomposition process of the cPINN/XPINN methods and its implicit mapping on the modern heterogeneous computing platform, the training time of the network can be reduced to a great extent. Recently, another domain decomposition method applied to the variational formulation of PINN is proposed by Kharazmi \etal\cite{hpVPINN} and named it \textit{hp-VPINN} method. 

\begin{figure}
\centering
\subfloat[] {
\includegraphics[ scale=0.39]{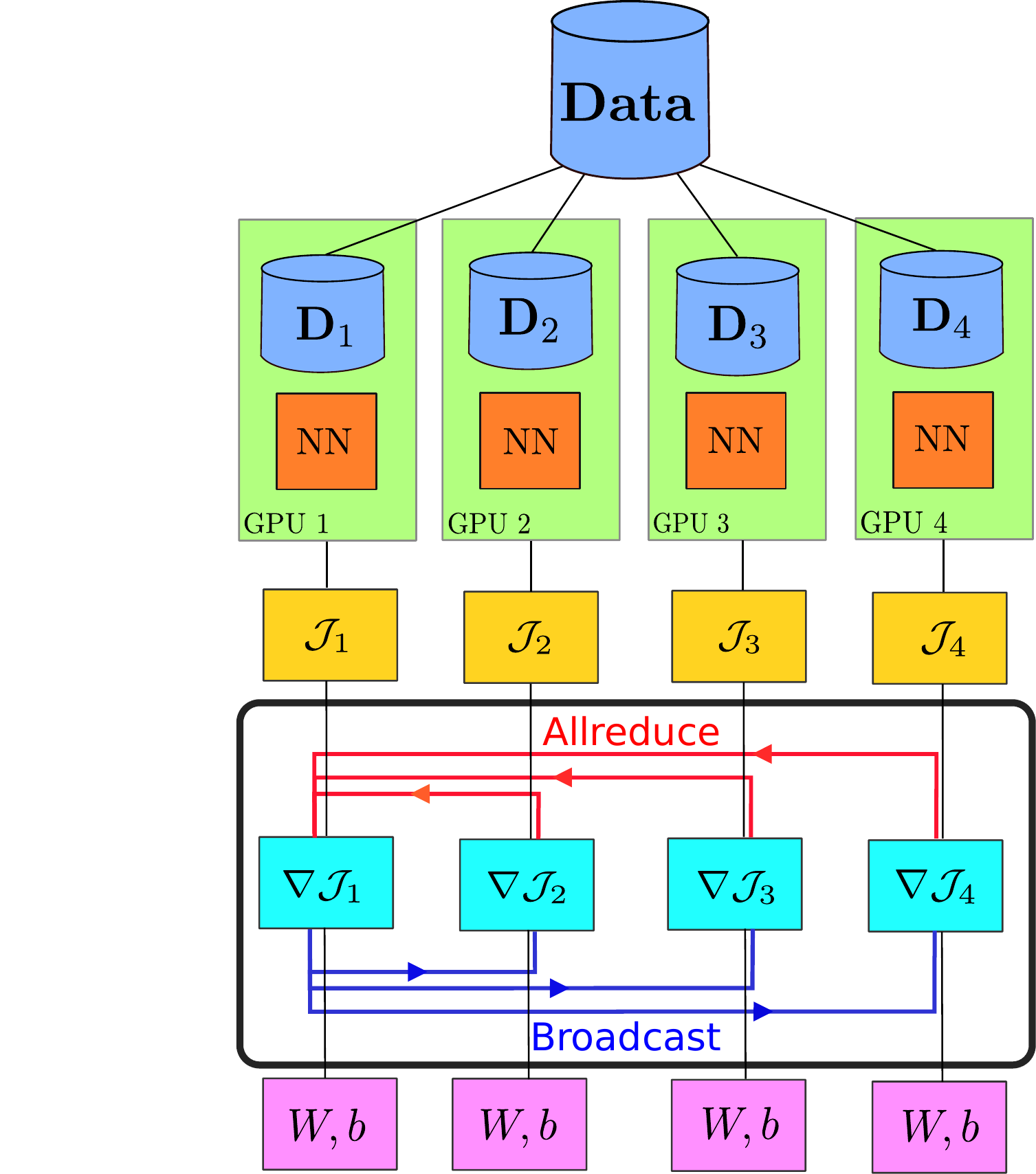}
\label{dp}
} \ \ \ \ \ \ \ \ 
\subfloat[]{
\includegraphics[ scale=0.76]{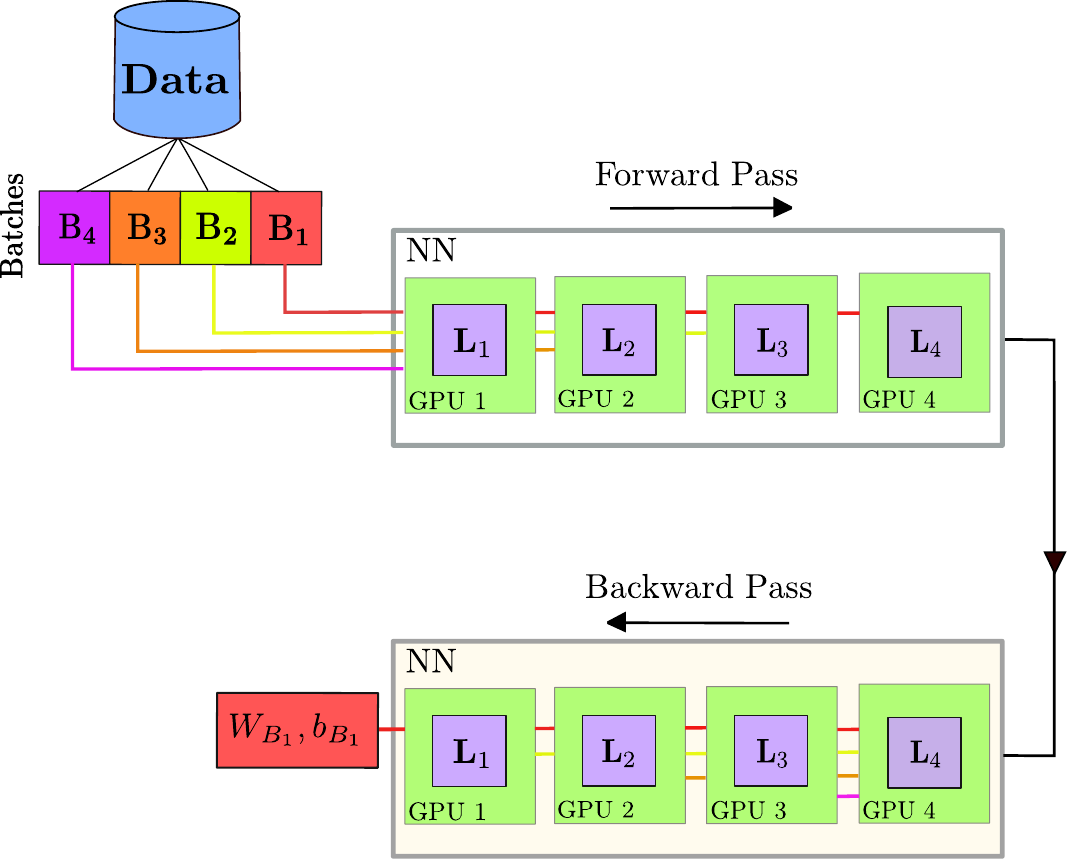}
\label{mp}
}
\caption{Schematic of the implementation of data and model parallel algorithms. (a) Method for the data-parallel approach, where the same neural network model, represented by NN, is loaded by each processor but works on a different chunk of input data. Synchronization of training (gradient of loss) is performed after the computation of loss on each processor via "allreduce" and "broadcast" operations represented by horizontal red and blue lines. (b) represents the model parallel approach, where each layer of the model (represented by $\text{L}_1 \ldots \text{L}_4)$ is loaded on a processor and each processor works on a batch of data $(\text{B}_1 \ldots \text{B}_4)$. Forward and backward passes are performed by using a pipeline approach. }
\label{data_model}
\end{figure}

There are currently two existing approaches for distributed training of neural networks, namely, the data-parallel approach \cite{sergeev2018horovod} and the model parallel approach. The data-parallel approach is based on the single instruction and multiple data (SIMD) parallel programming model, which results in a simple performance model driven by weak scaling. A block diagram showing the basic building blocks of the data-parallel approach is shown in Figure \ref{dp} for a four processors or co-processors system. The programming model used for the data-parallel approach falls into the regime of MPI+X system, where X is chosen as CPU(s) or GPU(s), depending on the size of the input data. In the data-parallel approach, the data is uniformly split into a number of chunks $(\text{D}_1,\ldots, \text{D}_4~\text{in Figure \ref{dp}})$, equal to the number of processors. The neural network (NN) model is initialized with the same parameters on all the processes as shown in Figure \ref{dp}. These neural networks are working on different chunks of the data, and therefore, work on different loss functions as shown by $\mathcal{J}_1, \ldots, \mathcal{J}_4$ in Figure \ref{dp}. 
In closely related work, NVIDIA introduced the parallel code SimNet \cite{hennigh2020nvidia}, which implements the standard PINN-based multi-physics simulation framework. The underlying idea of SimNet is to solve the differential equation by modeling the mass balance condition as a hard constraint as well as a global constraint. SimNet provides the functionality for multi-GPU and multi-Node implementation based on the data-parallel approach (Figure \ref{dp}).

To ensure a consistent neural network model (defined with same weights and biases) across all the processes and during each epoch or iteration of training, a distributed optimizer is used, which averages out the gradient of loss values $(\nabla \mathcal{J}_1,\ldots,\nabla \mathcal{J}_4~\text{in Figure \ref{dp} })$ on all the root processor through an "allreduce" operation. Subsequently, the averaged gradient of the loss function is sent to all the processors in the communicator world through a broadcast operation (collective communication). Then, the parameters of the model are updated through an optimizer. An additional component, which arises in the data-parallel approach, is the increase in global batch size with the number of processes. Goyal \etal \cite{goyal2017accurate} addressed this issue by multiplying the learning rate with the number of processes in the communicator. We note that in the data-parallel approach the model size (size of neural network parameters) remains uniform on each processor or GPU and that imposes a problem for large models to be trained on GPUs as they have fixed memory. 

To circumvent this issue, another distributed algorithm approach namely, the model parallel approach is proposed. 
A block diagram of the algorithm in the model parallel approach is shown in Figure \ref{mp}, which can be interpreted as a classic example of pipeline algorithm \cite{DeepSpeed}. In the model parallel approach, data is first divided into small batches $(\textbf{B}_1,\ldots,\textbf{B}_4~\text{in Figure \ref{mp}})$ and each hidden layer $(\textbf{L}_1,\ldots,\textbf{L}_4~\text{in Figure \ref{mp}})$ is distributed to a processor (or GPU). During the forward pass, once the $\textbf{B}_1$ is processed by $\textbf{L}_1$, the output of $\textbf{L}_1$ is passed as the input to $\textbf{L}_2$ and $\textbf{L}_1$ will start working on $\textbf{B}_2$ and so on. Once the $\textbf{L}_4$ finishes working, $\textbf{B}_1$ the backward pass (optimization process) will kickoff, thus, completing epochs of training in a pipeline mode. We note that the implementation of both algorithms are problem agnostic and do not incorporate any prior information on solutions to be approximated, which makes the performance of these algorithms to be dependent on the data size and model parameters. In the literature, the implementation of data and model parallel approaches are primarily carried out for problems pertaining to the classification and natural language processing \cite{goyal2017accurate, rasley2020deepspeed}, which are based on large amounts of training data. Therefore, the efficiency of data and the model parallel approach for scientific machine learning is not explored, which is primarily dominated by the high-dimensional and sparse data set. Apart from these two classical approaches, recently Xu \etal. \cite{xu2020distributed} deploy the topological concurrency on data structures of neural network parameters. In brief, this implementation could be comprehended as task-based parallelism and it is rooted in the idea of model-based parallelism. Additionally, it also provides interactivity with other discretization-based solvers such as Fenics \cite{alnaes2015fenics}.

The main advantage of cPINN and XPINN over data and model parallel approaches (including the work of Xu \etal \cite{xu2020distributed}) is their adaptivity for all hyperparameters of the neural network in each subdomain. In the vanilla data-parallel approach, the training across processors is synchronized by averaging the gradient of loss function across and subsequently broadcasting, enforcing the same network architectures on each processor. In scientific machine learning, the solution of the underlying physical laws is inferred based on a small amount of data, where the chosen neural network architecture depends on the complexity and regularity of the solutions to be recovered. To address this issue, the synchronization of the training process across all the processors has to be achieved by the physics of the problem, and therefore cPINNs and XPINNs come to the rescue. The convergence of the solution based on a single domain is constrained by its global approximation, which is relatively slow. On the other hand, computation of the solution based on local approximation by using the local neural networks can result in fast convergence. With regards to computational aspects, the domain decomposition-based approach requires a point-to-point communication protocol with a very small size of the buffer. However, for the data-parallel approach, it requires one \textit{allreduce} operation and one \textit{broadcast} operation with each having the buffer size equal to the number of parameters of the neural network, which makes communication across the processor slower. Moreover, for problems involving heterogeneous physics the data/model parallel approaches are not adequate, whereas cPINNs/XPINNs can easily handle such multi-physics problems.  Therefore, domain decomposition approaches paired with physics-based synchronization help more than the vanilla data-parallel approaches.

The main contributions of this paper are: 
\begin{itemize}
    \item We present a detailed unified distributed algorithm for cPINNs and XPINNs and implement it on CPU and GPU nodes. We also present weak scaling for CPUs and GPUs, and strong scaling for CPUs, demonstrating good speed-up and efficiency for both the methodologies on CPUs.    
    \item We present the performance of XPINN over cPINN for various forward problems such as nonlinear conservation laws and incompressible Navier-Stokes equations. We also implemented the XPINN method for solving an inverse problem involving a complex non-convex geometry with a non-trivial space-dependent conductivity, which is represented by a separate neural network. 
\end{itemize}

The paper is organized as follows. Section \ref{setup} defines the basic problem setup in detail. Section \ref{sec3} gives the mathematical setup for fully connected feed-forward neural networks. In section 4, the PINN algorithm, as well as cPINN and XPINN algorithms, are discussed in detail. Section 5 gives the details of the parallel implementation of cPINN and XPINN. In section 6, we discuss the optimization procedure on distributed systems. In section 7 we perform various computational experiments involving forward problems and one inverse problem. Finally, we summarize our findings in section 8.

\section{Problem setup} \label{setup}
The general form of a parametrized PDE is given by
\begin{align}\label{ProSet}
\mathcal{L}_{\mathbf{x}}(u;\lambda) & = f(\mathbf{x}),~~~~~ \mathbf{x} \in \Omega \subset \mathbb{R}^d,
\\ \mathcal{B}_k(u) & = g_k(\mathbf{x}),~~~~~ \mathbf{x} \in \Gamma_k \subset \partial \Omega \nonumber
\end{align}
for $k = 1, 2, \cdots, n_b$, where $\mathcal{L}_{\mathbf{x}}(\cdot)$ is the differential operator, $u$ is the solution, $\lambda = \{\lambda_1, \lambda_2,\cdots\}$ are the model parameters, $\mathcal{B}_k(\cdot)$ can be Dirichlet, Neumann, or mixed boundary conditions and $f(\mathbf{x})$ is the forcing term. Note that
for transient problems we consider time $t$ as one of the component of $\mathbf{x}$, and
the initial conditions can be simply treated as a particular type of boundary condition on the given spatio-temporal domain. The above setup encapsulates a wide range of problems in engineering and science. For equation \eqref{ProSet}, we define the residual $\mathcal{F}(u)$ as $
\mathcal{F}(u) \coloneqq \mathcal{L}_{\mathbf{x}}(u;\lambda) - f(\mathbf{x}).$ 

In this paper, we shall solve both forward problems, where the solution of a PDE
is inferred given fixed model parameters ($\lambda$) as well as inverse problems, where the unknown model parameters are learned from the observed data. The given mathematical model is converted into a surrogate model, more specifically, a given problem of solving PDE is converted into an optimization problem, where global minima of loss function correspond to the solution of the PDE. The loss function can be defined using training data points like initial and boundary conditions and the residual of the given PDE evaluated at random locations in the space-time domain.

\section{Fully connected feed-forward neural networks}\label{sec3}
Let $\mathcal{N}^L: \mathbb{R}^{D_i} \rightarrow \mathbb{R}^{D_o}$ be a feed-forward neural network of $L$ layers and $N_k$ neurons in $k^{th}$ layer ($N_0 = D_i$, and  $N_L = D_o$). The weight matrix and bias vector in the $k^{th}$ layer ($1 \leq k \leq L$) are denoted by $\bm{W}^k \in \mathbb{R}^{N_k \times N_{k-1}}$ and $\bm{b}^k \in \mathbb{R}^{N_k}$, respectively. The input vector is denoted by $\bm{z}\in \mathbb{R}^{D_i}$ and the output vector at $k^{th}$ layer is denoted by $\mathcal{N}^k(\bm{z})$ and $\mathcal{N}^0(\bm{z}) = \bm{z}$. We denote the activation function by $\Phi$, which is applied layer-wise along with the scalable parameters $n a^k$, where $n$ is the scaling factor. Layer-wise introduction of the additional parameters $a^k$ changes the slope of activation function in each hidden layer, thereby increasing the training speed. Moreover, these activation slopes can also contribute to the loss function through the slope recovery term, see \cite{jagtap2020adaptive, jagtap2020locally} for more details. 
Such locally adaptive activation functions enhance the learning capacity of the network, especially during the early training period.
Mathematically, one can prove this by comparing the gradient dynamics of the adaptive activation function method against that of the fixed activation method. The gradient dynamics of the adaptive activation modifies the standard dynamics (fixed activation) by multiplying a conditioning matrix by the gradient and by adding the approximate second-order term. In this paper, we used a scaling factor $n = 10$ for all hidden-layers and initialize $n a^k = 1, ~\forall k$, see~\cite{jagtap2020locally} for details.
The $(L-1)$-hidden layer feed-forward neural network is defined by
\begin{equation}
 \mathcal{N}^k(\bm{z}) = \bm{W}^k \Phi(a^{k-1} \mathcal{N}^{k-1}(\bm{z})) + \bm{b}^k \in \mathbb{R}^{N_k}, ~~~~2\leq k \leq L
\end{equation}
and $\mathcal{N}^1(\bm{z}) = \bm{W}^1 \mathcal{N}^0(\bm{z}) + \bm{b}^1$, where in the last layer, the activation function is identity.
By letting $\tilde{\boldsymbol{\Theta}} = \{ \bm{W}^k, \bm{b}^k, a^k \} \in \mathcal{V}$ as the collection of all weights, biases, and slopes, and taking $\mathcal{V}$ as the parameter space, we can write the output of the neural network as
$$u_{\tilde{\boldsymbol{\Theta}}}(\bm{z}) = \mathcal{N}^L(\bm{z}; \tilde{\boldsymbol{\Theta}}),$$
where $\mathcal{N}^L(\bm{z}; \tilde{\boldsymbol{\Theta}})$ emphasizes the dependence of the neural network output $\mathcal{N}^L(\bm{z})$ on $ \tilde{\boldsymbol{\Theta}}$. In general, weights and biases are initialized from known probability distributions.

\section{Brief overview of the PINNs, cPINNs and XPINNs}
\subsection{Physics-informed neural network (PINNs)}
Let $\{\mathbf{x}_u^{(i)}\}_{i=1}^{N_u}$ and $\{\mathbf{x}_F^{(i)}\}_{i=1}^{N_F}$ be the set of randomly selected training and residual points, respectively. These points are usually drawn from an unknown a priori distribution and chosen from a given input data. The PINN algorithm aims to learn a surrogate $ u \approx u_{\tilde{\boldsymbol{\Theta}}}$ to compute the solution $u$ for a given PDE.
The loss function for the PINN is given as 
\begin{align}
\label{p-loss}
\mathcal{L}\left(\tilde{\bm{\Theta}}\right) = W_u~ \text{MSE}_u \left(\tilde{\bm{\Theta}};\{\mathbf{x}_u^{(i)}\}_{i=1}^{N_u}\right) + W_\mathcal{F}~ \text{MSE}_\mathcal{F} \left(\tilde{\bm{\Theta}};\{\mathbf{x}_F^{(i)}\}_{i=1}^{{}N_F}\right),
\end{align}
where $W_u$ and $W_\mathcal{F}$ are the weights for the data and residual losses, respectively. The mean square error is given by
\begin{align*}
\text{MSE}_u \left(\tilde{\bm{\Theta}};\{\mathbf{x}_u^{(i)}\}_{i=1}^{N_u}\right)&=\myfrac{1}{N_u}\sum_{i=1}^{N_u} \left |u^{(i)} - u_{\tilde{\boldsymbol{\Theta}}} \left( \mathbf{x}_u^{(i)}\right) \right |^2,\\
\text{MSE}_\mathcal{F} \left(\tilde{\bm{\Theta}};\{\mathbf{x}_F^{(i)}\}_{i=1}^{N_F}\right)&=\myfrac{1}{N_F}\sum_{i=1}^{N_F} \left | \mathcal{F}_{\tilde{\boldsymbol{\Theta}}}\left(\mathbf{x}^{(i)}_F \right) \right|^2,
\end{align*}
where $\text{MSE}_u$ is the MSE for data mismatch term enforcing the given initial/boundary conditions as constraints, resulting in a well-posed problem. The term $\text{MSE}_\mathcal{F}$ is the MSE for PDE residual, with $\mathcal{F}_{\tilde{\boldsymbol{\Theta}}}=\mathcal{F}{(u_{\tilde{\boldsymbol{\Theta}}})}$ representing the residual of governing PDEs. The parameters of the neural networks $u_{\tilde{\boldsymbol{\Theta}}}$ are computed by minimizing the loss function in (\ref{p-loss}).

Both experimental, as well as synthetic training data, can be incorporated into the loss function. The PDE residual can be computed using automatic differentiation (AD) \cite{baydin2018automatic}. AD is an accurate way to calculate derivatives in a computational graph compared to numerical differentiation since they do not suffer
from errors such as truncation and round-off errors. Thus, evaluation of the PDE operator is achieved with such graph-based differentiation, which can
be incorporated in the loss function along with the training data. This way, PINN can be viewed as a grid-free approach, thus avoiding the tyranny of the mesh generation procedure.

\subsection{Domain decomposition approaches in physics-informed neural networks}
In this section, we will briefly describe the two recently proposed domain decomposition approaches in the PINN framework, namely, cPINN and XPINN. The computational domain is divided into $N_{sd}$ number of non-overlapping regular/irregular subdomains, and in each subdomain, we deploy the separate neural networks, which communicate with each other via a common interface. 
In the cPINN and XPINN frameworks, the output of the neural network in the $q^{th}$ subdomain is given by
$
u_{{\tilde{\boldsymbol{\Theta}}}_q}(\bm{z}) = \mathcal{N}^L(\bm{z}; \tilde{\boldsymbol{\Theta}}_q) \in \Omega_q, ~~~ q = 1,2, \ldots, N_{sd}.$
Then, the final solution is obtained as 
\begin{equation}\label{FSol}
 u_{\tilde{\bm{\Theta}}}(\bm{z}) = \sum_{q = 1}^{N_{sd}} u_{{\tilde{\bm{\Theta}}}_q}(\bm{z}) \cdot \mathbbm{1}_{\Omega_q}(\bm{z}),
 \end{equation}
where the indicator function $\mathbbm{1}_{\Omega_q}(\bm{z})$ is defined as
\begin{equation*}
 \mathbbm{1}_{\Omega_q}(\bm{z}) \coloneqq \begin{cases} 0 & \text{If}~ \bm{z} \notin \Omega_q, \\ 
                                                   1 & \text{If}~ \bm{z} \in \Omega_q \backslash \text{Common interface in the $q^{th}$ subdomain}, \\
                                                   \frac{1}{\mathcal{S}} & \text{If}~ \bm{z} \in \text{Common interface in the $q^{th}$ subdomain}, \\
                                     \end{cases}
\end{equation*}
where $\mathcal{S}$ represents the total number of subdomains intersecting along the common interface.

Let $\{\mathbf{x}^{(i)}_{u_q}\}_{i=1}^{N_{u_q}} $, $\{\mathbf{x}^{(i)}_{F_q}\}_{i=1}^{N_{F_q}} $ and $\{\mathbf{x}^{(i)}_{I_q}\}_{i=1}^{N_{I_q}}$ be the set of randomly selected training, residual, and the common interface points, respectively in the $q^{th}$ subdomain.  $N_{u_q}, N_{F_q}$, and $ N_{I_q}$ represent the number of training data points, the number of residual points, and the number of points on the common interface in the $q^{th}$ subdomain, respectively.
Similar to PINN, the cPINN and XPINN methods aim to learn a surrogate $u_q \approx u_{\tilde{\boldsymbol{\Theta}}_q}$, $q = 1,2,\cdots, N_{sd}$ for predicting the solution $u \approx u_{\tilde{\boldsymbol{\Theta}}}$ of the given PDE over the entire computational domain using equation \eqref{FSol}. The loss function of cPINN/XPINN is defined subdomain-wise, which has a similar structure as the PINN loss function in each subdomain but it is endowed with the interface conditions for stitching the subdomains together. 

\subsubsection{Conservative PINNs (cPINNs) \cite{jagtap2020conservative}}
In cPINN method, the loss function in the $q^{th}$ subdomain is defined as
\begin{align}\label{cPINNLoss}
 \mathcal{J}(\tilde{\boldsymbol{\Theta}}_q)  = ~ & W_{u_q} ~ \text{MSE}_{u_q}(\tilde{\boldsymbol{\Theta}}_q; \{\mathbf{x}^{(i)}_{u_q}\}_{i=1}^{N_{u_q}} ) + W_{\mathcal{F}_q} ~ \text{MSE}_{\mathcal{F}_q}(\tilde{\boldsymbol{\Theta}}_q;\{\mathbf{x}^{(i)}_{F_q}\}_{i=1}^{N_{F_q}} ) + W_{I_q} ~ \underbrace{   \text{MSE}_{u_{avg}}(\tilde{\boldsymbol{\Theta}}_q;\{\mathbf{x}^{(i)}_{I_q}\}_{i=1}^{N_{I_q}} )}_{\text{Interface condition}} \\ \nonumber
  & + W_{I_{\text{flux}_q}} ~ \underbrace{ \text{MSE}_{\text{flux}}(\tilde{\boldsymbol{\Theta}}_q;\{\mathbf{x}^{(i)}_{I_q}\}_{i=1}^{N_{I_q}} )}_{\text{Interface condition}},
\end{align}
where $~ q = 1,2,\cdots, N_{sd}.$ 
The MSE is given for each term by
\begin{align*}
\text{MSE}_{u_q}(\tilde{\boldsymbol{\Theta}}_q;\{\mathbf{x}^{(i)}_{u_q}\}_{i=1}^{N_{u_q}} ) & = \frac{1}{N_{u_q}} \sum_{i=1}^{N_{u_q}}\left|u^{(i)}_q - u_{{\tilde{\boldsymbol{\Theta}}}_q}(\mathbf{x}^{(i)}_{u_q})\right|^2, \ \\  \text{MSE}_{\mathcal{F}_q}(\tilde{\boldsymbol{\Theta}}_q;\{\mathbf{x}^{(i)}_{F_q}\}_{i=1}^{N_{F_q}} )  &= \frac{1}{N_{F_q}} \sum_{i=1}^{N_{F_q}}\left|\mathcal{F}_{{\tilde{\boldsymbol{\Theta}}}_q}(\mathbf{x}^{(i)}_{F_q})\right|^2,
\\ \text{MSE}_{u_{avg}}(\tilde{\boldsymbol{\Theta}}_q;\{\mathbf{x}^{(i)}_{I_q}\}_{i=1}^{N_{I_q}} )  &= \sum_{\forall q^+} \left( \frac{1}{N_{I_q}} \sum_{i=1}^{N_{I_q}}\left| u_{{\tilde{\boldsymbol{\Theta}}}_q}(\mathbf{x}^{(i)}_{I_q})-  \left\{\!\left\{u_{{\tilde{\boldsymbol{\Theta}}}_q}(\mathbf{x}^{(i)}_{I_q})\right\}\!\right\}\right|^2 \right),
\\ \text{MSE}_{\text{flux}}(\tilde{\boldsymbol{\Theta}}_q;\{\mathbf{x}^{(i)}_{I_q}\}_{i=1}^{N_{I_q}} ) & = \sum_{\forall q^+} \left( \frac{1}{N_{I_q}} \sum_{i=1}^{N_{I_q}}\left|  f_{{\tilde{\boldsymbol{\Theta}}}_q} (\mathbf{x}^{(i)}_{I_q}) \cdot \bm{n} - f_{{\tilde{\boldsymbol{\Theta}}}_{q^+}}(\mathbf{x}^{(i)}_{I_q})\cdot \bm{n}\right|^2 \right).
\end{align*}
The $W_{u_q}, W_{\mathcal{F}_q}, W_{I_{\text{flux}_q}}$ and $W_{I_q}$ are the data mismatch, residual, and interface (both, normal flux as well as average solution continuity along the interface) weights, respectively.  The term $\text{MSE}_{\text{flux}}$ represents the normal flux continuity term along the interface. The terms $\text{MSE}_{u_q} $ and $\text{MSE}_{\mathcal{F}_q}$ are the same as before described in the PINN algorithm, and  $\mathcal{F}_{{\tilde{\boldsymbol{\Theta}}}_q} \coloneqq \mathcal{F}(u_{{\tilde{\boldsymbol{\Theta}}}_q})$ represent the residual of the governing PDEs in the $q^{th}$ subdomain. The average value of $u$ along the common interface is given by $\left\{\!\left\{u_{{\tilde{\boldsymbol{\Theta}}}_q} \right\}\!\right\} = u_{avg} \coloneqq \frac{u_{{\tilde{\boldsymbol{\Theta}}}_q} + u_{{\tilde{\boldsymbol{\Theta}}}_{q^+}}}{2}$ (assuming that along the common interface only two subdomains intersect). The interface conditions ensure that the information from the one subdomain can be propagated throughout the neighboring subdomains. These conditions also play an important role in the convergence of subdomains, where no training data are available.

\subsubsection{Extended PINNs (XPINNs) \cite{jagtap2020extended}}
In XPINN method, along with the average solution continuity condition, the more general residual continuity conditions are also imposed along the interface, which allows solving any differential equations with space-time decomposition. Apart from these continuity conditions, additional continuity conditions like normal flux continuity, higher-order solution derivative continuity, integral constraints (invariants) continuity conditions, etc. can be imposed depending on the orientation of the interface and the nature of the governing differential equations. In XPINN method, the loss function in the $q^{th}$ subdomain is defined as
\begin{align}\label{XPINNLoss}
 \mathcal{J}(\tilde{\boldsymbol{\Theta}}_q)  = ~ & W_{u_q} ~ \text{MSE}_{u_q}(\tilde{\boldsymbol{\Theta}}_q; \{\mathbf{x}^{(i)}_{u_q}\}_{i=1}^{N_{u_q}} ) + W_{\mathcal{F}_q} ~ \text{MSE}_{\mathcal{F}_q}(\tilde{\boldsymbol{\Theta}}_q;\{\mathbf{x}^{(i)}_{F_q}\}_{i=1}^{N_{F_q}} ) + W_{I_q} ~ \underbrace{   \text{MSE}_{u_{avg}}(\tilde{\boldsymbol{\Theta}}_q;\{\mathbf{x}^{(i)}_{I_q}\}_{i=1}^{N_{I_q}} )}_{\text{Interface condition}} \\ \nonumber
  & + W_{I_{\mathcal{F}_q}} ~ \underbrace{ \text{MSE}_{\mathcal{R}}(\tilde{\boldsymbol{\Theta}}_q;\{\mathbf{x}^{(i)}_{I_q}\}_{i=1}^{N_{I_q}} )}_{\text{Interface condition}}  + \underbrace{\text{Additional Interface Condition's}}_{\text{Optional}},
\end{align}
where $~ q = 1,2,\cdots, N_{sd}.$ The $W_{I_{\mathcal{F}_q}}$ is the weight for the residual continuity condition. 
Again, the MSE is given as
\begin{align*}
\text{MSE}_{u_q}(\tilde{\boldsymbol{\Theta}}_q;\{\mathbf{x}^{(i)}_{u_q}\}_{i=1}^{N_{u_q}} ) & = \frac{1}{N_{u_q}} \sum_{i=1}^{N_{u_q}}\left|u^{(i)}_q - u_{{\tilde{\boldsymbol{\Theta}}}_q}(\mathbf{x}^{(i)}_{u_q})\right|^2, \ \\  \text{MSE}_{\mathcal{F}_q}(\tilde{\boldsymbol{\Theta}}_q;\{\mathbf{x}^{(i)}_{F_q}\}_{i=1}^{N_{F_q}} )  &= \frac{1}{N_{F_q}} \sum_{i=1}^{N_{F_q}}\left|\mathcal{F}_{{\tilde{\boldsymbol{\Theta}}}_q}(\mathbf{x}^{(i)}_{F_q})\right|^2,
\\ \text{MSE}_{u_{avg}}(\tilde{\boldsymbol{\Theta}}_q;\{\mathbf{x}^{(i)}_{I_q}\}_{i=1}^{N_{I_q}} )  &= \sum_{\forall q^+} \left( \frac{1}{N_{I_q}} \sum_{i=1}^{N_{I_q}}\left| u_{{\tilde{\boldsymbol{\Theta}}}_q}(\mathbf{x}^{(i)}_{I_q})-  \left\{\!\left\{u_{{\tilde{\boldsymbol{\Theta}}}_q}(\mathbf{x}^{(i)}_{I_q})\right\}\!\right\}\right|^2 \right),
\\ \text{MSE}_{\mathcal{R}}(\tilde{\boldsymbol{\Theta}}_q;\{\mathbf{x}^{(i)}_{I_q}\}_{i=1}^{N_{I_q}} ) & = \sum_{\forall q^+} \left( \frac{1}{N_{I_q}} \sum_{i=1}^{N_{I_q}}\left|  \mathcal{F}_{{\tilde{\boldsymbol{\Theta}}}_q} (\mathbf{x}^{(i)}_{I_q}) - \mathcal{F}_{{\tilde{\boldsymbol{\Theta}}}_{q^+}}(\mathbf{x}^{(i)}_{I_q})\right|^2 \right),
\end{align*}
 The $\text{MSE}_{\mathcal{R}}$ is the residual continuity condition on the common interface given by two different neural networks on subdomains $q$ and $q^+$, respectively; the superscript $+$ over $q$ represents the neighboring subdomain(s). Both $\text{MSE}_{\mathcal{R}}$ and $\text{MSE}_{u_{avg}}$ terms are defined for all neighbouring subdomain(s) $q^+$, which is shown in their respective expressions by the summation sign over all $q^+$.

\section{Parallel implementation of cPINN and XPINN}
\begin{figure}
\centering
 \includegraphics[ scale=0.6]{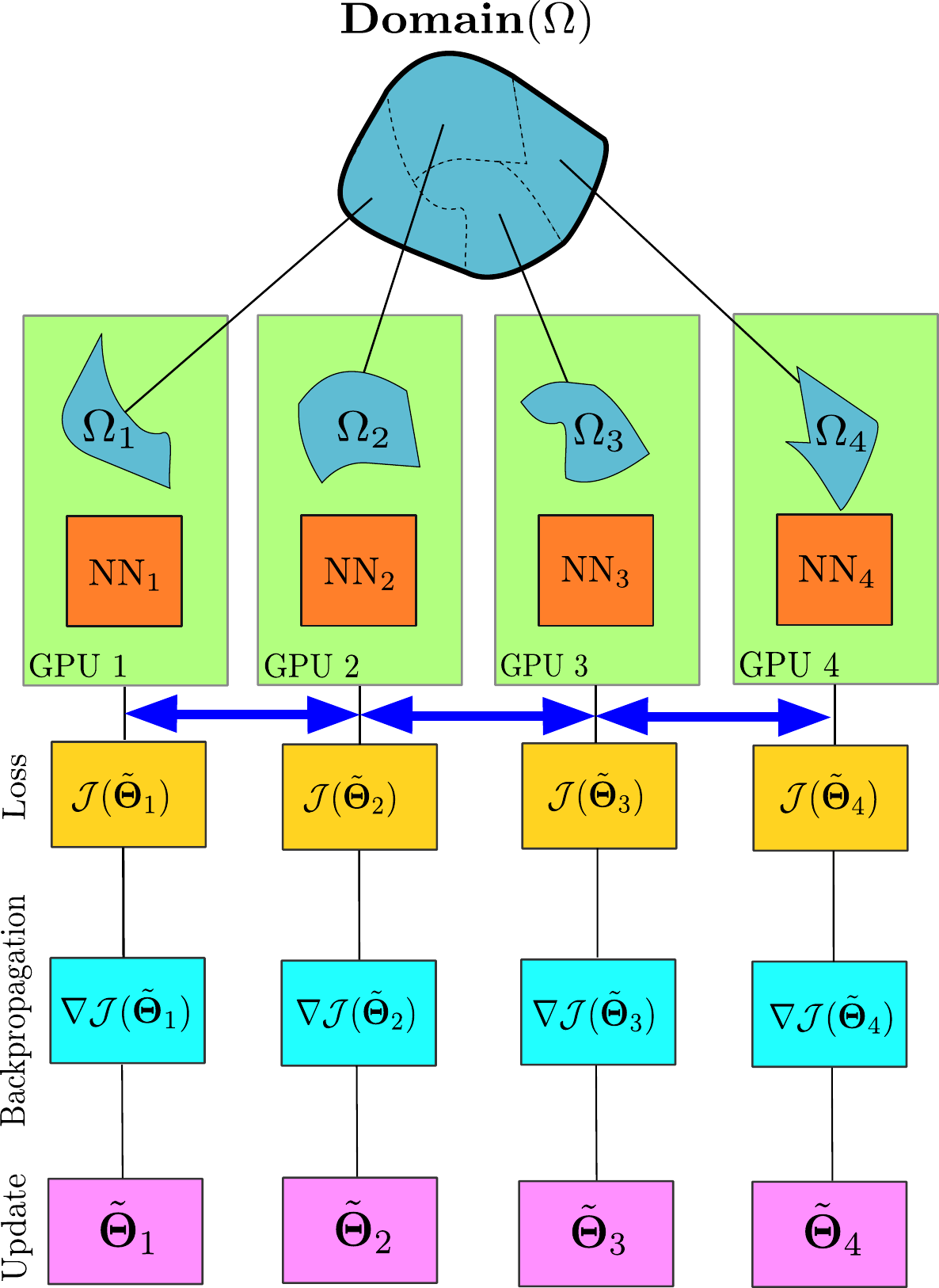}
\caption{Building blocks for distributed cPINN and XPINN methodologies deployed on a heterogeneous computing platform. The domain $\Omega$ is decomposed into several subdomains $\Omega_1,\ldots, \Omega_4$ equal to the number of processors, and individual neural networks (NN$_1, \ldots,$ NN$_4$) are employed in each subdomain, which gives separate loss functions $\mathcal{J}(\tilde{\boldsymbol{\Theta}}_q), q = 1,\ldots,4$ coupled through the interface conditions (shown by the double-headed blue arrow). The trainable parameters are updated by finding the gradient of loss function individually for each network and penalizing the
continuity of interface conditions.}
   \label{domain-decomp0}
\end{figure}

In this section, we describe the algorithm implemented for parallelizing the cPINN and XPINN methods. Before diving into the finer level of implementation, we will first explain the primary building blocks of cPINN and XPINN on a distributed heterogeneous (CPU's and GPU's) computing platform in the purview of traditional data and model parallel approach. Figure \ref{domain-decomp0} shows a detailed schematic of cPINN and XPINN algorithms. At first, the entire domain $\Omega$ (as shown in Figure \ref{domain-decomp0}) is decomposed into several subdomains $(\Omega_1,\ldots, \Omega_4)$ equal to the number of processors or co-processors (GPU) and each subdomain will be assigned to a separate neural network model. Each processor will compute their local solution in the respective subdomain, which is stitched along the non-overlapping interface using the corresponding interface conditions just before the computation of loss (shown by the double-headed blue arrow in Figure \ref{domain-decomp0}). Thus, cPINN and XPINN inherit the advantages of both data and model parallel approaches, in addition to the imposition of prior information on the solution, which eventually determines the parameters of neural network in each subdomain. 

Next, we will describe the systematic implementation of cPINN and XPINN in a distributed and heterogeneous computing environment. We have split up the entire implementation into two stages, (i) pre-processing, and (ii) parallel solution stages. A detailed implementation level is given by Algorithm \ref{algo}, which will be explained further in subsequent sections. 

\subsection{Pre-processing}
\begin{figure}
\centering
 \includegraphics[trim=3cm 3.5cm 3cm 0.5cm,clip, scale=0.6]{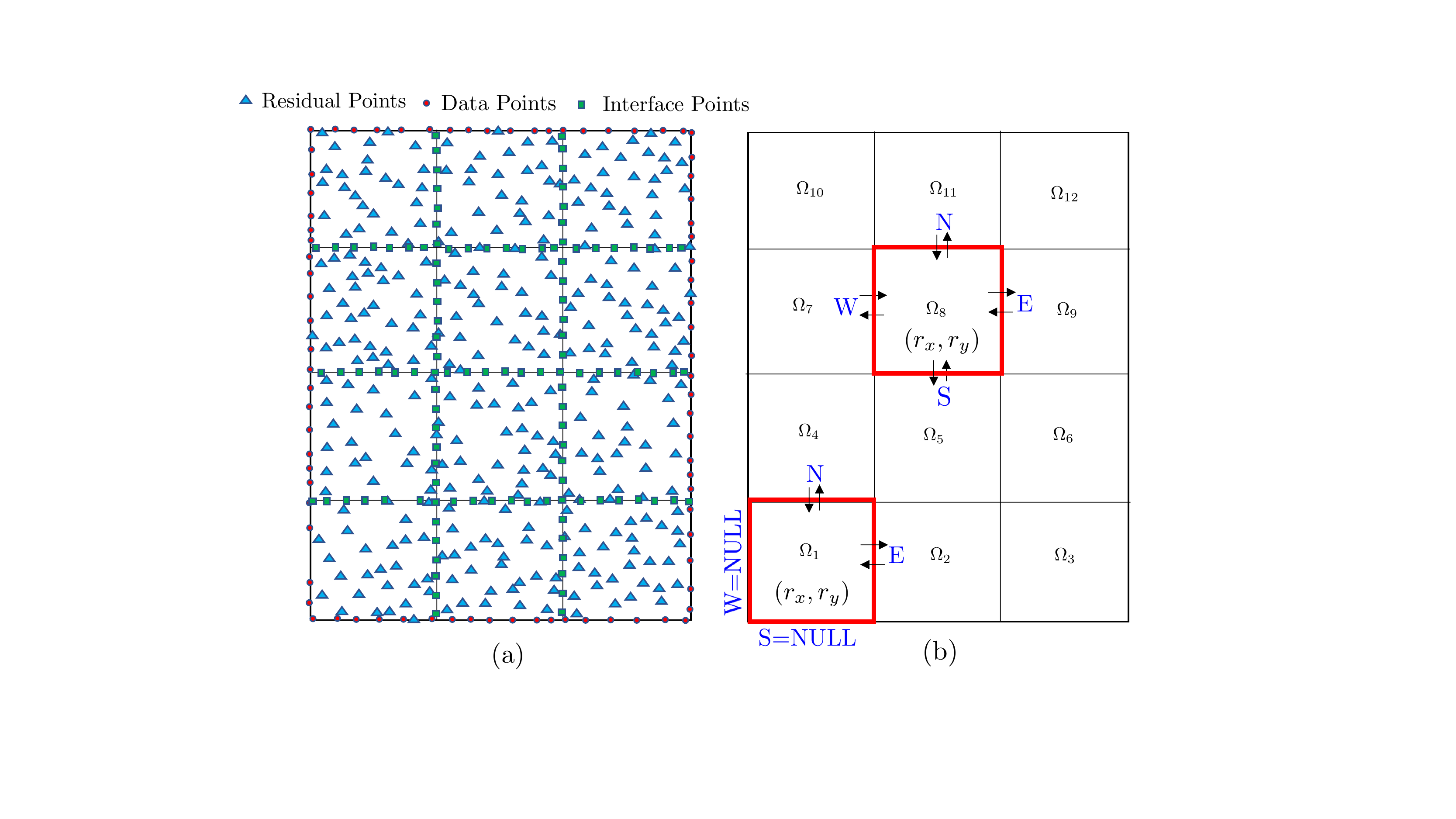}
\caption{ A domain decomposition approach for cPINN and XPINN. (a) 12 subdomains with locations of residual (blue triangles), data (red circles), and interface (green squares) points;  (b) two-way point-to-point communication between subdomains with their neighbors. In particular, the communication for subdomains $\Omega_1$ and $\Omega_8$ involves two and four edges, respectively, as shown.}
   \label{domain-decomp}
\end{figure}

The pre-processing stage primarily deals with the decomposition of the computational domain $\Omega$, shown in a blue color font in Algorithm \ref{algo}, and preparing a data structure for concurrent access. The method of domain decomposition is widely used in numerical simulation for solving PDEs related to the problems of fluid flow \cite{tang2020review}, wave propagation, and heat flow \cite{dolean2015introduction}. The implementation of a parallel algorithm for cPINN and XPINN is partially inspired by these conventional domain decomposition methods. In these methods, we first divide the domain into non-overlapping subdomains such that $\Omega=\bigcup_{q=1}^{N_{sd}} \Omega_q$, with a constraint, that the total number of $N_{sd}$ equals the number of processors or accelerators to be used for the computation, as shown in Figure \ref{domain-decomp}a. For example, in Figure \ref{domain-decomp} the domain is divided into 12 subdomains, and for ease of presentation, the subdomains are created using Cartesian topology. However, it is important to note that the implementation of cPINN/XPINN methods is not mandatory for the Cartesian topology and can be extended to any irregularly shaped geometry. In each subdomain, the training, residual and interface points are sampled by using the \textit{i.i.d} approach drawn from $~\mathcal{N} (0, \sigma^2)$. To store and access the training data concurrently, we use a dictionary type data structure with key-value pairs being the subdomain number and list containing the indices of edges of the subdomains conforming with the boundary of the domain and represent it with EToV. This data structure is further utilized during the solver stage explained later. Thereafter, we prepared another dictionary data structure containing the indices of subdomains and a 2D array of $(x, y)$ coordinates of residual points as key-value pair, respectively. Finally, we prepare another dictionary data structure to store the residual points along the edges of the subdomains. The edges corresponding to the interior subdomains are extracted by subtracting the sets corresponding to the global edge number with edges conforming with the boundary (EToV). Finally, in the pre-processing stage, we will endow each processor, corresponding to each subdomain in a one-to-one way,  with physical coordinates $(x, y)$ of the residual points in respective subdomains (blue triangles in Figure \ref{domain-decomp}a), interface points along the edges of the subdomain (green square in Figure \ref{domain-decomp}a), and training data mapped to respective subdomains (red color circles in Figure \ref{domain-decomp}a). Although this is explained for the two-dimensional domain, the same approach can be easily extended to higher-dimensional domains because, the data and residuals are represented in an arbitrary fashion unlike the conventional domain decomposition, where points have to be stored in an ordered fashion.  

\subsection{Parallel solution}
This stage is further divided into two stages: first, computation and second, communication stage. In the computation stage during each epoch, $u_{{\tilde{\boldsymbol{\Theta}}}_q}(\mathbf{x}_{I_q})$, $\mathcal{F}_{{\tilde{\boldsymbol{\Theta}}}_q} (\mathbf{x}_{I_q})$ and $f_{{\tilde{\boldsymbol{\Theta}}}_q} (\mathbf{x}_{I_q})$ are computed in each subdomain concurrently as computation of these terms does not require any data from neighboring elements. This part of the algorithm is represented in red color font in Algorithm \ref{algo}.

During the communication stage, represented by the green-colored font in Algorithm \ref{algo}, we first arrange the processors in a logical 2D layout, as shown in Figure \ref{domain-decomp}b with each processor represented by there $x$-and $y~$- rank expressed as
\begin{align}\label{rmap}
(r_x, r_y) = (r//N_x,~ r\%N_y),
\end{align}
where $(r_x, r_y)$ represent local ranks of processor,  $r$ is the global rank of processor, and $N_x$ and $N_y$ are the number of subdomains in $x$ and $y$ directions. The literals ``//" and `\%`" represent the integer division and modulo operations, respectively. 
The local rank of each processor will assist in determining the local rank of neighboring subdomains, represented by \textit{South, East, North, West} $(\text{S},\text{E}, \text{N}, \text{W})$ edges. The local ranks of neighboring subdomains can be computed by increasing or decreasing the values of $r_x$ and $r_y$, depending on the direction of neighbors. For example, the neighboring subdomains of $\Omega_8$ (Figure \ref{domain-decomp}b)  in the south, east, north, and west direction is expressed as $(r_x-1, r_y),~(r_x, r_y + 1),~(r_x + 1, r_y),~\text{and}~(r_x, r_y-1)$, respectively. The $r_x$ and $r_y$ have to be non-negative, and negative $r_x$ and $r_y$ correspond to boundary edges and therefore the local ranks are initialized as NULL. In MPI implementation, communication is carried out by using the global rank, which is obtained by mapping the local rank to the global rank using $r=r_x*N_x + r_y$. Subdomains aligning with the boundary domain will have two neighbors for a 2D layout (see subdomain $\Omega_1$ in Figure \ref{domain-decomp}b) and remaining non-existing subdomains are initialized as NULL with MPI construct of \text{MPI.PROC\_NULL}.
Now, as the preparation of the message envelope for communication is ready,  $u_{{\tilde{\boldsymbol{\Theta}}}_q}(\mathbf{x}_{I_q})$ and $f_{{\tilde{\boldsymbol{\Theta}}}_q} (\mathbf{x}_{I_q})$ are sent and received by each processor by using the point-to-point \textit{send-receive} protocol in a \textit{non-blocking} way \cite{gropp2014using}.
To receive the data from neighboring subdomains we initialize a buffer corresponding to each edge of subdomains. Once the communication is completed and the processors are synchronized the subdomain-wise loss is computed using (\ref{cPINNLoss}) or (\ref{XPINNLoss}). Thereafter, optimizations of loss functions corresponding to each subdomain are carried out concurrently and parameters for each neural network are retrieved. We note that the processors are mapped to GPUs or accelerators using the same approach as mentioned above but using CUDA-aware MPI \cite{lonvcar2016openmp}.

\begin{algorithm}
\algsetup{linenosize=\tiny}
\textbf{Step 0}: Initialize MPI, rank and size of processors $ = N_{sd}$\\
set method = 0 for cPINN\\
set method = 1 for XPINN \\
comm=MPI.Init()\\
rank=MPI.Get\_rank(comm)\\
size=MPI.Get\_size(comm)\\
\textcolor{blue}{
\If{rank == 0 }{
Divide the computational domain into $N_{sd}$ number of non-overlapping regular subdomains\\
Specify the training data, residual and interface points in all subdomains $\Omega_q$\\
\textit{Training data} : $u_{\tilde{\boldsymbol{\Theta}}_q}$ network $\{\mathbf{x}^{(i)}_{u_q}\}_{i=1}^{N_{u_q}},~~~q = 1,2,\cdots, N_{sd}.$ \\
\textit{Residual training points} :  $\mathcal{F}_{{\tilde{\boldsymbol{\Theta}}}_q}$ network $\{\mathbf{x}^{(i)}_{F_q}\}_{i=1}^{N_{F_q}},~~~q = 1,2,\cdots, N_{sd}$.\\
 \textit{Interface points} : $\{\mathbf{x}^{(i)}_{I_{q}}\}_{i=1}^{N_{I_q}},~~~q = 1,2,\cdots, N_{sd}$ with $\{\mathbf{x}^{(i)}_{I_{q}}\}_{i=1}^{N_{I_q}} \in \{S,~E,~N,~W\}, \text{which represents South, East, North and West edges for all subdomains}.$}
}

\textcolor{red}{
\For{epoch in Nepochs}{
 $u_{\tilde{\boldsymbol{\Theta}}_q}$=DNN($\{\mathbf{x}^{(i)}_{u_q}\}_{i=1}^{N_{u_q}}$, $\mathbf{W}_q$, $\mathbf{b}_q$);\\
 compute $\mathcal{F}_{\tilde{\boldsymbol{\Theta}}_q}  \left( \mathbf{x}_{F_q}\right)$,\\
 compute $u_{{\tilde{\boldsymbol{\Theta}}}_q}(\mathbf{x}_{I_q})$\\
 \If{method == 0}{
 compute flux $f_{\tilde{\boldsymbol{\Theta}}_q}(\mathbf{x}_{I_q})$\\
 }\Else{
 compute residual $\mathcal{F}_{\tilde{\boldsymbol{\Theta}}_{q}} \left(\mathbf{x}_{I_q} \right)$ 
 }
 {\color{green!45!black}
 \For{ i in (0, size)}{
 \If {i == rank} {
 \If {method == 0}{
 MPI.Isend($ u_{{\tilde{\boldsymbol{\Theta}}}_q}(\mathbf{x}_{I_q})$, r), where $r \in \{S_r, E_r, N_r, W_r\}$ with $S_r,E_r, N_r, W_r$ being the neighboring processes\\
 MPI.Isend($f_{\tilde{\boldsymbol{\Theta}}_q}(\mathbf{x}_{I_q})$, r)\\
  MPI.Irecv($ u_{{\tilde{\boldsymbol{\Theta}}}_q}(\mathbf{x}_{I_q})$, r)\\
 MPI.Irecv($f_{\tilde{\boldsymbol{\Theta}}_q}(\mathbf{x}_{I_q})$, r)\\
 MPI.waitall()\\
 compute $\mathcal{J}\left( \tilde{\boldsymbol{\Theta}}_q \right)$
 }
 \Else{
 MPI.Isend($ u_{{\tilde{\boldsymbol{\Theta}}}_q}(\mathbf{x}_{I_q})$, r), where $r \in \{S_r, E_r, N_r, W_r\}$ with $S_r,E_r, N_r, W_r$ being the neighboring processes\\
 MPI.Isend($\mathcal{F}_{\tilde{\boldsymbol{\Theta}}_{q}} \left(\mathbf{x}_{I_q} \right)$, r)\\
  MPI.Irecv($ u_{{\tilde{\boldsymbol{\Theta}}}_q}(\mathbf{x}_{I_q})$, r)\\
 MPI.Irecv($\mathcal{F}_{\tilde{\boldsymbol{\Theta}}_{q}} \left(\mathbf{x}_{I_q} \right)$, r)\\
 MPI.waitall()\\
 compute $\mathcal{J}\left( \tilde{\boldsymbol{\Theta}}_q \right)$
 }
 }
 }
 }
Compute the model parameters by minimizing the loss function using an optimization process in each subdomain independently\\
 $\tilde{\boldsymbol{\Theta}}_q^* = \text{arg~min}_{{\tilde{\boldsymbol{\Theta}}}_q \in \mathcal{V}_q} \mathcal{J}(\tilde{\boldsymbol{\Theta}}_q),$
 here $\mathcal{V}_q$ is the parameter space in the $q^{th}$ subdomain.
}
}
\caption{Parallel algorithm for cPINN and XPINN; Textblocks in {\color{blue}blue}, {\color{red}red}, and {\color{green!45!black}green} colors represent the preprocessing, computation and communication stages, respectively. (For interpretation of the references to color, the reader is referred to the web version of this article.)}
\label{algo}
\end{algorithm}

\section{Optimization on distributed systems}
We seek to find the optimal parameters $\tilde{\boldsymbol{\Theta}}_q^*$
that minimize the loss function $\mathcal{J} (\tilde{\boldsymbol{\Theta}}_q)$ in each subdomain.  Although there is no theoretical guarantee that the optimization process converges to a global minimum, our computational experiments indicate that as long as the given PDE is well-posed and has a unique solution, the cPINN/XPINN formulations are capable of achieving an accurate solution provided that a sufficiently expressive network and a sufficient number
of residual points are used. There are several optimization algorithms available to minimize the loss function. In general, a gradient-based optimization method is employed
for the training of parameters. 
The stochastic gradient descent (SGD) method is the widely used optimization
method. In SGD, a small set of points are randomly sampled to find the direction of
the gradient in every iteration. The SGD algorithm works well to avoid bad local minima
during training of DNN under one point convexity property. A brief survey on SGD is
given by Ruder \cite{ruder2016overview}. 

The optimization of single loss functions on the distributed systems can be performed without significant change in the code. In the literature, for single loss function-based methodologies many distributed optimization procedures are proposed.  Le \etal \cite{le2011optimization} used a distributed optimization procedure for classification problems. In \cite{dean2012large}, Dean \etal. proposed the \textit{Sandblaster} framework that supports a variety of distributed batch optimization procedures, including a distributed implementation of L-BFGS \cite{byrd1995limited} method. In the present work, both cPINN and XPINN methods involve multiple loss functions, which need to be optimized in parallel. In the present work, we used an ADAM optimizer for minimizing the loss functions in each subdomain.

\section{Computational experiments}
\label{ce}
In this section, we first define the terminology about the scaling and the performance of the distributed algorithm. Then, we present the profiling of the vanilla PINN as a pedagogical example, which establishes a computational perspective of the PINN methodology. In the rest of the section, we test and discuss the accuracy and performance of the Algorithm \ref{algo} for forward and inverse problems.
To generate the performance results, we used IBM's Watson Machine Learning package (IBM$\_$WML), which include TensorFlow 1.15 and installed on Red Hat Enterprise Linux (RHEL) version 7.6. To code up the algorithm, we have used Python 3.6. The versions of CUDA and CUDA drivers are 10.2 and 440.33.01, respectively.

\subsection{Weak scaling}
In the weak scaling metric, the assignment of work on each processor remains constant, and as the problem size increases the number of processing elements also increases, which results in a constant workload per processor. This type of scaling is used for a problem that does not fit in the memory of a single node and is commonly known as a memory-bound problem. The weak scaling efficiency, $W_e$, is expressed as
\begin{equation}\label{wse}
W_e = \myfrac{T_1}{T_{N_{PW}}},
\end{equation}
where $T_1$ is the time required to complete the work with one processor and $T_{N_{PW}}$ is the time taken by $N_{PW}$ processor to complete the $N_{PW}$ unit of the same work with $N_{PW}$ processing elements. 

\subsection{Strong scaling}
In the strong scaling metric, the problem size is fixed and the number of processing elements is increased, which results in a reduced workload per processor. These measurements are employed for problems that are compute-bound. The strong scaling efficiency is expressed as
\begin{equation}\label{sse}
S_e = \myfrac{s}{N_{P}},
\end{equation}
where $s$ is the speed up  defined as $s=T_1/T_{{N_P}}$, with $T_{{N_P}}$ being the time taken by $N_P$ processors to complete the same amount of work.

\subsection{A pedagogical example on the computational complexity of PINNs}
Before diving into the parallel implementation of cPINN and XPINN, we first describe the profiling of PINN and its dependence on various hyperparameters. In particular, the depth, width of the network, and the number of residual points, which can significantly affect the performance of the neural network. To this end, we consider the one-dimensional time-dependent Burgers equation given by
\begin{equation}\label{BE}
u_t + u u_x - (0.01/\pi) u_{xx}=0, ~~~ x\in [-1, 1],~t>0,
\end{equation}
with initial and boundary conditions $u(0, x) = -\sin(\pi x)$ and  $u(t, -1) = u(t,1)=0$, respectively.
In all cases, we used hyperbolic tangent activation function and the learning rate is 1e-4.

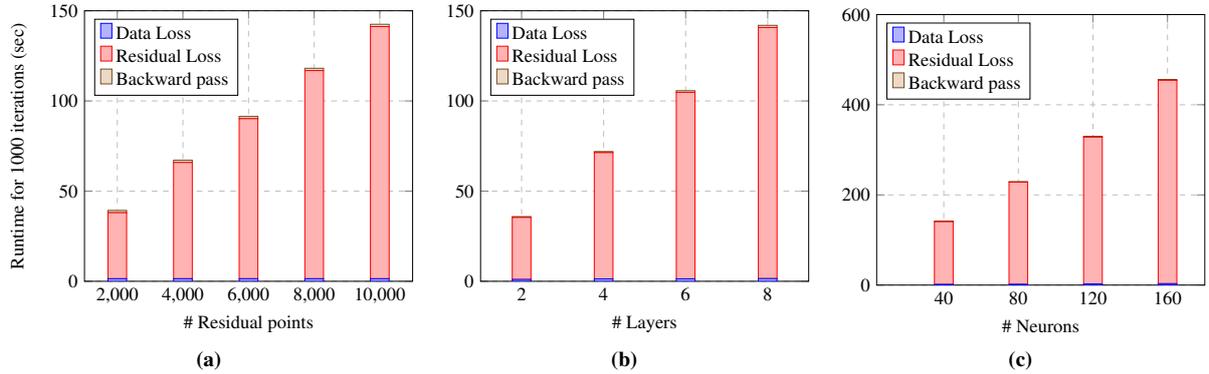
\begin{figure}
\centering
\subfloat[]{\begin{tikzpicture} [scale=0.70] \label{pvf}
\begin{axis}[
	scale=0.9,
	legend cell align=left,
	xlabel={$\#$ Residual points},
	ylabel={Runtime for 1000 iterations (sec)},
	xmin=1000, xmax=11000,
	ymin=0,ymax=150,
	ybar stacked,
	xtick={2000, 4000, 6000, 8000, 10000},
	xticklabel style={
        /pgf/number format/fixed,
     },
    scaled x ticks=false,
	legend pos=north west,
	xmajorgrids=true,
	ymajorgrids=true,
	grid style=dashed,
] 
\addplot coordinates{(2000, 1.42)  (4000,  1.42) (6000, 1.42) (8000, 1.42) (10000, 1.42)};
\addplot coordinates{(2000, 36.6096) (4000, 64.4694) (6000, 88.7513) (8000, 115.4202) (10000, 139.8164)};
\addplot coordinates{(2000, 1.2724) (4000, 1.2724) (6000, 1.2724) (8000, 1.2724) (10000, 1.2724)};
\legend{Data Loss, Residual Loss, Backward pass}
\end{axis}
\end{tikzpicture}} \ \ \ \ 
\subfloat[]
{\begin{tikzpicture}[scale=0.70]\label{pvl}
\begin{axis}[
	scale=0.9,
	legend cell align=left,
	xlabel={$\#$ Layers},
	ylabel={},
	xmin=1, xmax=9,
	ymin=0,ymax=150,
	ybar stacked,
	xtick={2, 4, 6, 8},
	xticklabel style={
        /pgf/number format/fixed,
     },
    scaled x ticks=false,
	legend pos=north west,
	xmajorgrids=true,
	ymajorgrids=true,
	grid style=dashed,
] 
\addplot coordinates{(2, 1.0051)  (4,  1.2965) (6, 1.3415) (8, 1.4867)};
\addplot coordinates{(2, 34.4016) (4, 70.0465) (6, 103.4207) (8, 139.2269)};
\addplot coordinates{(2, 0.4654) (4, 0.5993) (6, 0.9309) (8, 1.2078)};
\legend{Data Loss, Residual Loss, Backward pass}
\end{axis}
\end{tikzpicture}}\ \ \ \ 
\subfloat[]
{\begin{tikzpicture}[scale=0.70]\label{pvn}
\begin{axis}[
	scale=0.9,
	legend cell align=left,
	xlabel={$\#$ Neurons},
	ylabel={},
	xmin=4, xmax=180,
	ymin=0,ymax=600,
	ybar stacked,
	xtick={40, 80, 120, 160},
	xticklabel style={
        /pgf/number format/fixed,
     },
    scaled x ticks=false,
	legend pos=north west,
	xmajorgrids=true,
	ymajorgrids=true,
	grid style=dashed,
] 
\addplot coordinates{(40, 1.4867)  (80,  1.7549) (120, 2.6158) (160, 3.1509)};
\addplot coordinates{(40, 139.2269) (80, 226.4963) (120, 325.6283) (160, 451.0887)};
\addplot coordinates{(40, 1.2078) (80, 1.2842) (120, 1.5608) (160, 1.622)};
\legend{Data Loss, Residual Loss, Backward pass}
\end{axis}
\end{tikzpicture}}
\caption{Cost profile of the PINN algorithm for the one-dimensional Burgers equation: The computation time is represented for data loss, residual loss, and backward pass, separately; (a) runtime for 1000 iterations against the number of residual points with 200 data points; (b) runtime against the number of hidden-layers with fixed 200 data points and 10000 residual points, and 40 neurons per layer; (c) runtime against the number of neurons in each layer with fixed 200 data points and 10000 residual points, with 8 hidden-layers.}
\end{figure}

Figure 4 shows the plot of compute times, consisting of the data loss, residual loss, and backward pass. In particular,  Figure \ref{pvf} shows the computational cost for varying numbers of residual points. In this case, the computation is performed for 200 data points, 8 hidden-layers with 40 neurons in each layer.  Figure \ref{pvf} shows that the maximum time is consumed by the computation of residual loss, which involves the computation of partial derivatives in the governing PDEs. The partial derivative is computed by using the method of reverse mode auto-differentiation \cite{baydin2018automatic}, which involves the traversal of the computation graph. The computational complexity of graph traversal depends on the order of underlying PDE(s), the number of hidden-layers, and the number of neurons in each layer. Figure \ref{pvf} also shows that with the increase in the number of residual points the computation time for residual loss increases. Figure \ref{pvl} represents the plot of computation time of PINN against the depth of the network. The computation is performed for 200 data points, 10000 residual points, and 40 neurons in each layer.  As the depth of the network increases the computation time for the residual loss increases. Figure \ref{pvn} represents the variation of compute times with respect to the number of neurons in each layer. The computation is performed using 200 data points and 8 hidden-layers.  There is a positive correlation between compute time and width of neural network with maximum time spent on residual loss. The computation reported in Figure 4 is performed on \textit{Intel's Cascade Lake} processors with 32-bit floating points.

\subsection{Two-dimensional steady-state incompressible  Navier–Stokes  equations}
The governing steady-state  incompressible Navier-Stokes equation in two-dimensions are given as
\begin{align} \label{NSEq}
\begin{aligned}
\mathbf{u} \cdot \nabla \mathbf{u} &=-\nabla p + \frac{1}{Re} \nabla^2 \mathbf{u}, \qquad \text{in}~\Omega\\
\nabla \cdot \mathbf{u} &= 0, \qquad \text{in}~\Omega
\end{aligned}
\end{align}
with appropriate Dirichlet boundary conditions, see \cite{jagtap2020conservative} for more details.
The $\mathbf{u}=(u,v), p$ and $Re$ represents the velocity components (in $x-$ and $y-$ directions), pressure and Reynolds number, respectively. The neural network used in each subdomain consists of 5 hidden-layers with 80 neurons in each layer. The activation function is hyperbolic tangent and the learning rate is 6e-4.

\subsubsection{Accuracy of Algorithm \ref{algo}}
First, we test the accuracy of our parallel algorithm presented in Algorithm \ref{algo}. To measure the accuracy, we solve the steady-state two-dimensional problem of incompressible Navier–Stokes equations to compute the solution for the problem of the lid-driven cavity. 
We compute the solution of (\ref{NSEq}) for $Re=100$,
with $\Omega=[0,1] \times [0, 1]$ and no-slip boundary conditions are imposed on the wall boundary. To compute the solution of (\ref{NSEq}) using cPINN and XPINN, we decomposed $\Omega$ in four subdomains with each subdomain endowed with 12000 residual points, 80 boundary points, and 250 interface points. A total of 4 QuadroRTX 6000 GPUs, each with 24 GB memory is used for computation. The inter-connection between GPUs is achieved through PCI Express and communication between GPUs is employed via CUDA aware MPI. Figure 5 (top row) shows the contour plots of velocity components $u$ and $v$ obtained by using the parallel cPINN method. A comparison between the solutions obtained from cPINN, XPINN and the reference solutions by Ghia \etal \cite{ghia} are represented in Figure 5 (bottom row). The inferred solutions show a very good agreement with the reference solution and therefore confirm the accuracy of the distributed cPINN and XPINN algorithms (Algorithm \ref{algo}).
\begin{figure}
	\centering
	\includegraphics[trim={1cm, 2cm, 2cm, 1cm}, clip, scale=0.34]{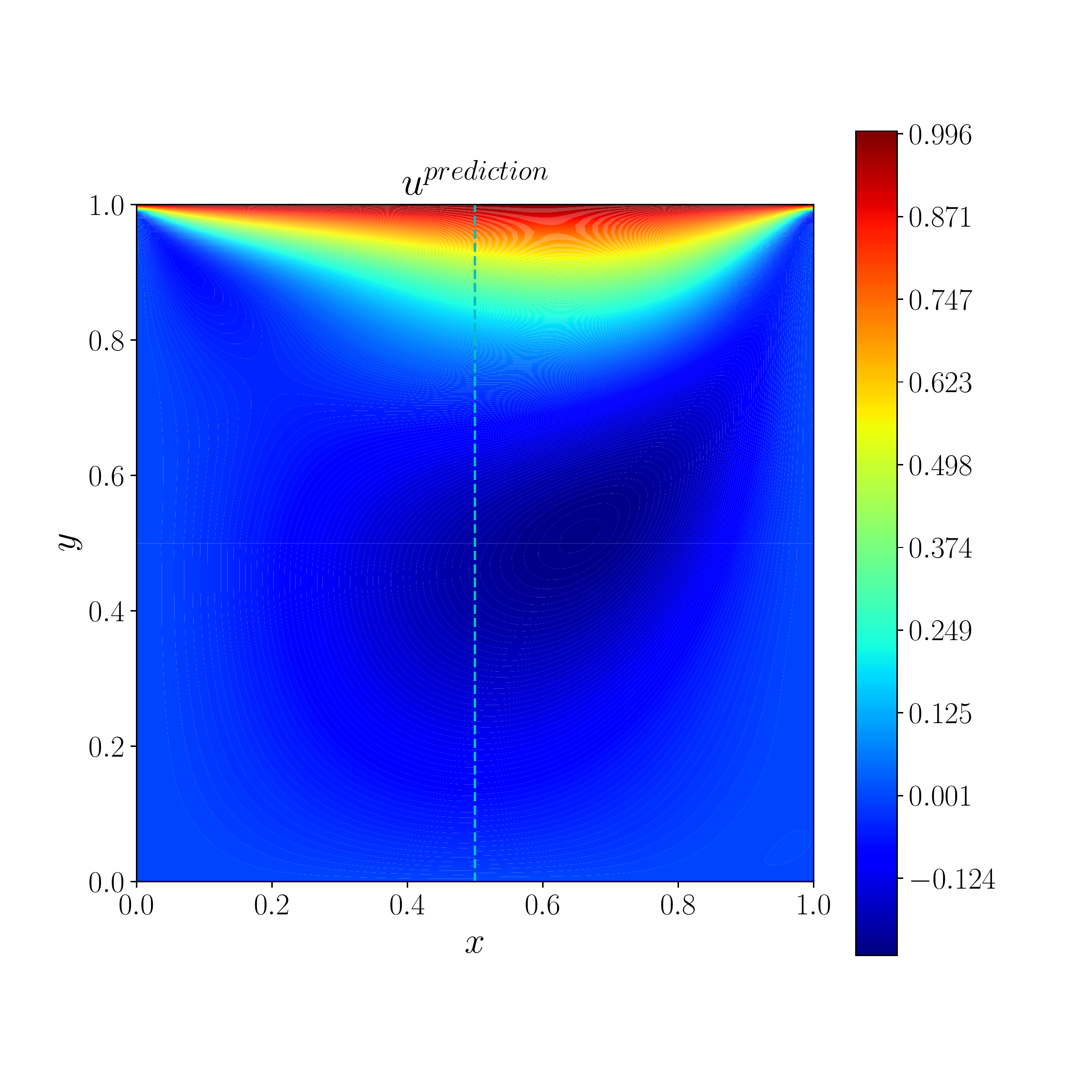}
	\includegraphics[trim={1cm, 2cm, 2cm, 1cm}, clip, scale=0.34]{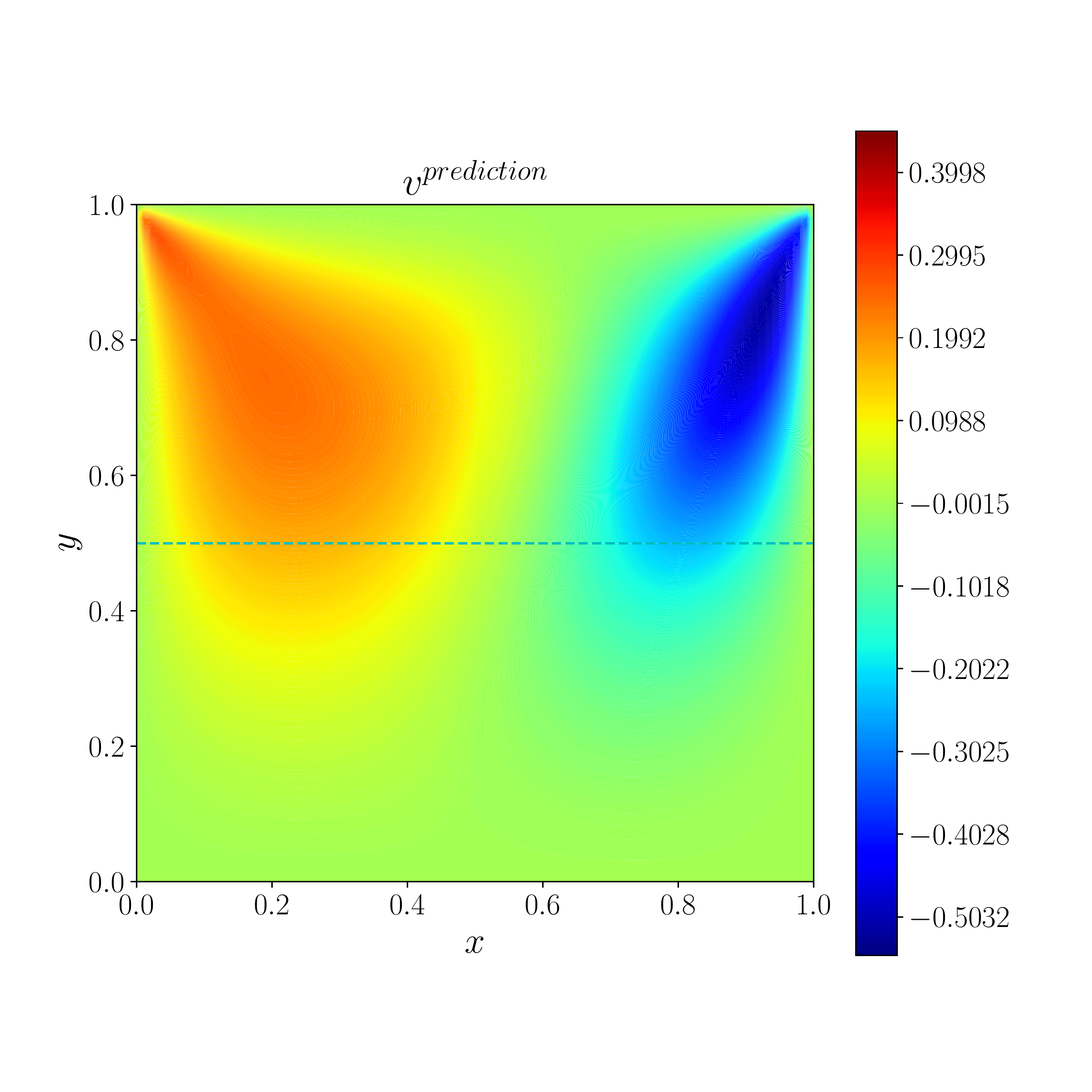}
	\includegraphics[trim={0cm, 0cm, 0cm, 0cm}, clip, scale=0.385]{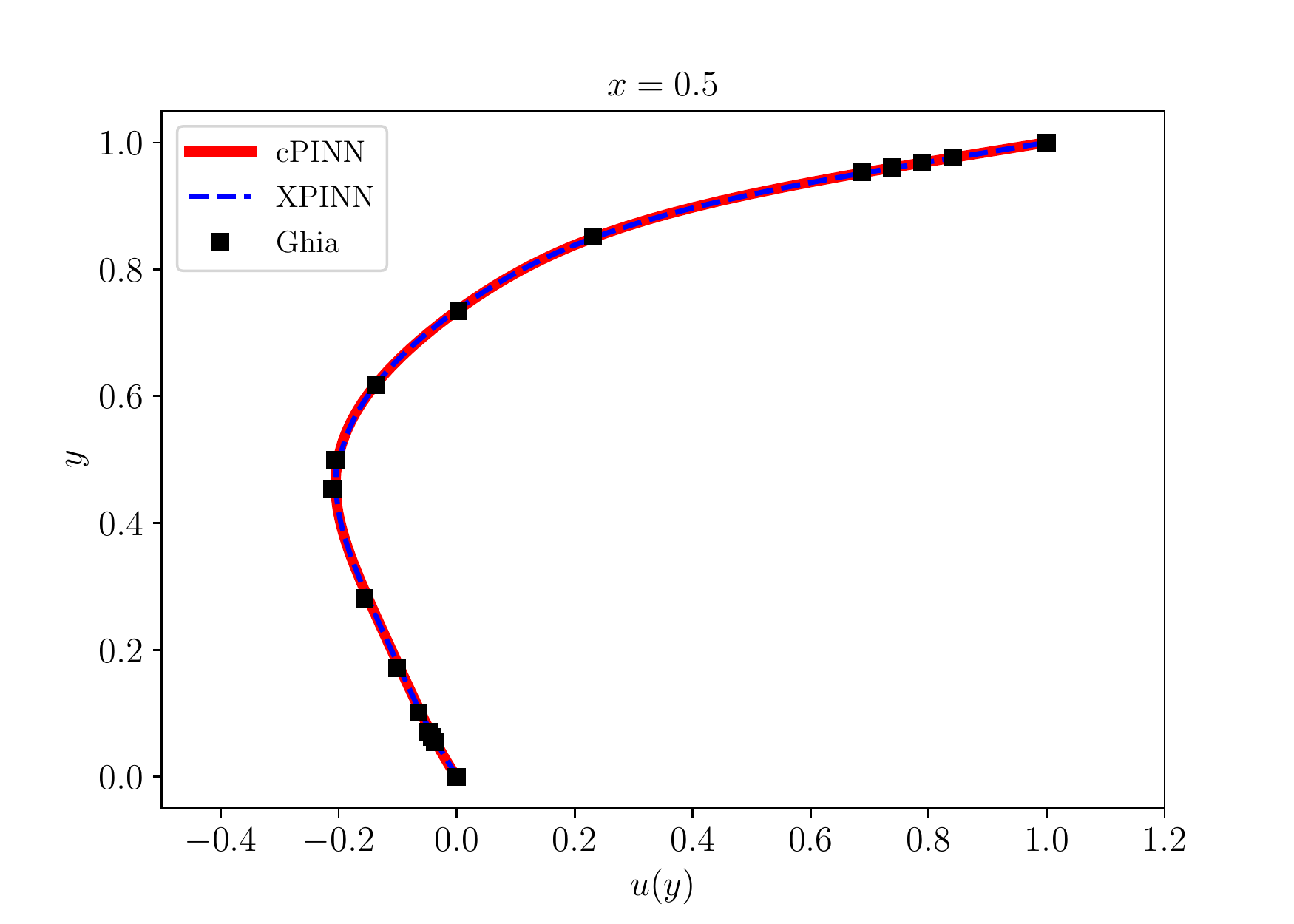}
	\includegraphics[trim={0cm, 0cm, 0cm, 0cm}, clip, scale=0.385]{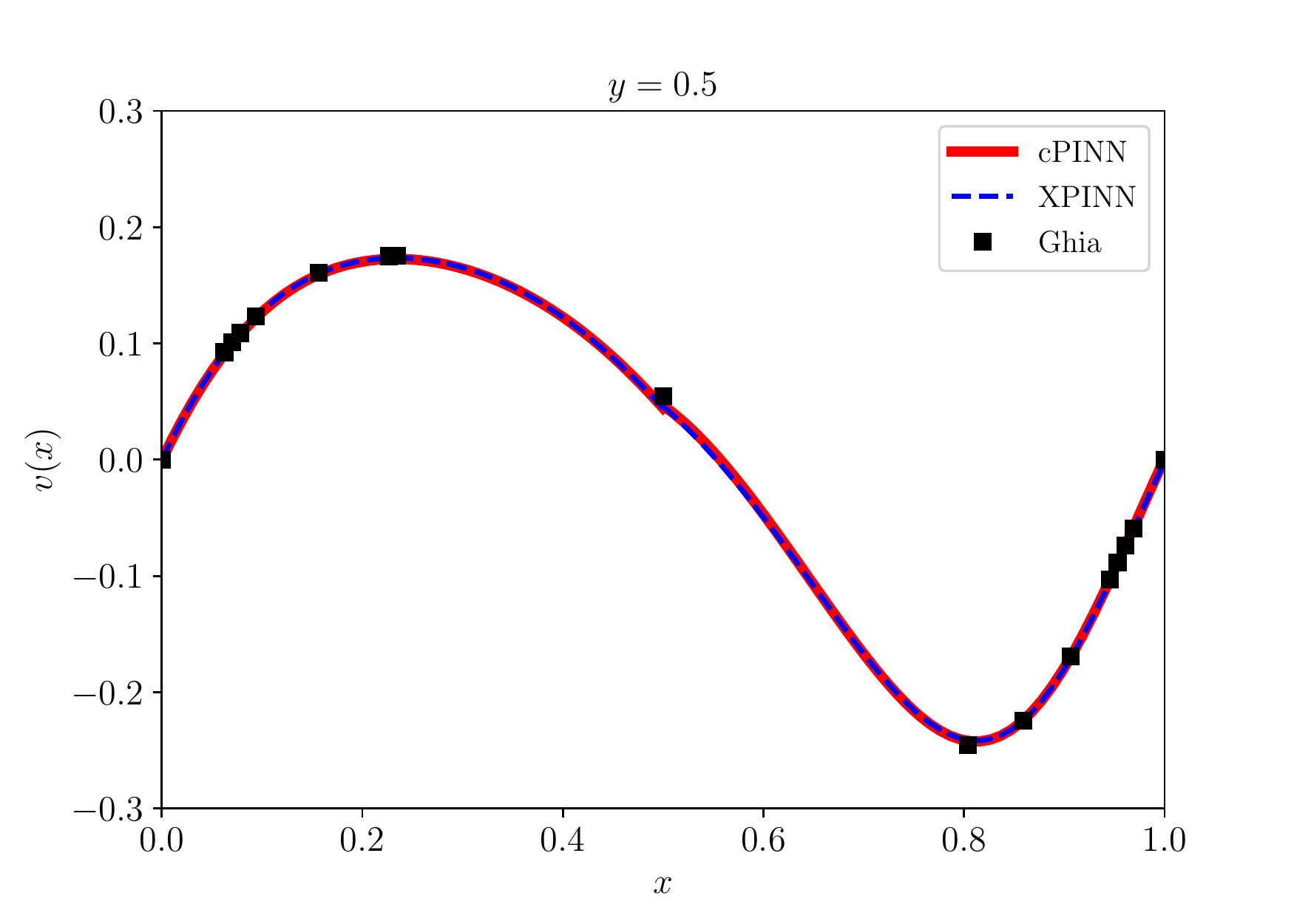}
	\caption{Two-dimensional incompressible Navier-Stokes equations: The top row shows the velocity contour plots for cPINN and the bottom row shows the comparison of cPINN and XPINN velocities with the results obtained by Ghia \etal \cite{ghia} for Re = 100.}
\end{figure}

\subsubsection{Computation \textit{versus} Communication times}
\begin{figure}
\centering
\subfloat[]{\begin{tikzpicture}\label{cc_cpinn}
\begin{axis}[
	scale=0.9,
	legend cell align=left,
	xlabel={Nodes},
	ylabel={Runtime for 10 iterations (sec)},
	xmin=2, xmax=26,
	ymin=0,ymax=14,
	ybar stacked,
	xtick={4,8, 12, 16, 20, 24},
	legend pos=north west,
	xmajorgrids=true,
	ymajorgrids=true,
	grid style=dashed,
] 
\addplot coordinates{(4,2.0) (8,2.1) (12, 2.0) (16, 2) (20, 2.0) (24, 2.0)};
\addplot+[
        error bars/.cd,
            y dir=both, 
            y explicit,
            error bar style={line width=1.0pt, color=black},
    ]coordinates{
         (4,3.3)+-(0, 0.17) 
         (8, 4.1)+-(0, 0.0599)
         (12, 3.7)+-(0, 0.12)
         (16, 3.5)+-(0, 0.1019)
         (20,3.57)+-(0,0.1411)
        (24, 3.7)+-(0,0.174)
};
\legend{Computation Time, Communication Time}
\end{axis}
\end{tikzpicture}} \ \ \ \ \ \ 
\subfloat[]{
\begin{tikzpicture}\label{cc_xpinn}
\begin{axis}[
    scale=0.9,
	legend cell align=left,
	xlabel={Nodes},
	ylabel={},
	xmin=2, xmax=26,
	ymin=0,ymax=15,
	ybar stacked,
	xtick={4,8, 12, 16, 20, 24},
	legend pos=north west,
	xmajorgrids=true,
	ymajorgrids=true,
	grid style=dashed,
] 
\addplot coordinates{(4,4.0) (8,3.0) (12,2.7) (16, 2.9) (20, 3.0) (24, 3.0)};
\addplot+[
        error bars/.cd,
            y dir=both, 
            y explicit,
            error bar style={line width=1.0pt, color=black},
    ]coordinates{
         (4,6.4)+-(0, 0.60) 
         (8, 7.9)+-(0, 0.1143)
         (12, 7.3)+-(0, 0.64)
         (16, 7.1)+-(0, 0.3956)
         (20, 7.0)+-(0,0.1577)
        (24, 7.0)+-(0,0.1018)
};
\legend{Computation Time, Communication Time}
\end{axis}
\end{tikzpicture}} 
\caption{Two-dimensional incompressible Navier-Stokes equations: Computation and communication time using the CPU implementation
for (a) cPINN and (b) XPINN, with 100 residual points and 20 interface points.}
\label{cc}
\end{figure}
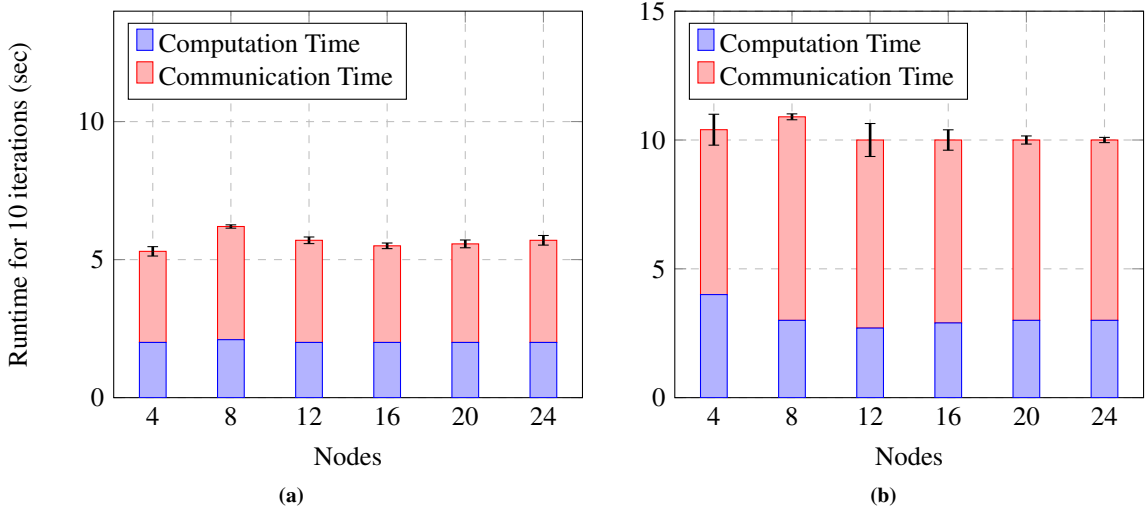

Now, we show runtimes for the distributed cPINN and XPINN algorithms spent during the computation and communication phases. Figures \ref{cc_cpinn} and \ref{cc_xpinn} represent the bar plots for computation and communication times corresponding to cPINN and XPINN, respectively. The domain is decomposed from 4 to 24 subdomains, where 200 residual and 20 interface points are sampled randomly in each subdomain. A small number of interface points is chosen so that the problem becomes more communication dominated, and performance is measured in a weak scaling fashion. The run-times in Figure \ref{cc} show the plot of computation and communication time for 10 iterations of training, measured for \textit{Intel's cascade lake} CPUs, against the number of nodes with each rank mapped on one node. Figure 6a shows that computation and communication times contribute almost equally to the total time with a perfect weak scaling. However, for XPINN in Figure \ref{cc_xpinn}, the communication time is more than the computation time and more importantly than the cPINN implementation, which is explained later in the context of the problem described by (\ref{NSEq}). Further, we perform the same exercise by adding one GPU with each CPU, with 1 CPU + 1 GPU corresponding to each rank and per node, and results are presented in Figure \ref{cc_gpu}. The CPU and GPUs on each node are connected with NVLink and nodes are connected with a topology of non-blocking fat-tree using Mellanox's EDR as a fabric. Each node contains IBM POWER9 CPUs and NVIDIA's Volta V100 GPUs. Figures \ref{cp_gpu} and \ref{xp_gpu} represent the computation and communication time for cPINN and XPINN, respectively. In Figure \ref{cc_gpu}, the domain is also decomposed from 4 to 24 subdomains where 4000 residual and 200 interface points are sampled randomly in each subdomain. The interpretation of Figure \ref{cc_gpu} is similar to the results of Figure \ref{cc}. Figure \ref{xp_gpu} also shows that communication in XPINN is greater than cPINN (Figure \ref{cp_gpu}) but both show a perfect weak scaling.

\begin{table}
\caption{Fluxes for the steady-state two-dimensional incompressible 
Navier-Stokes equations.}
\label{flux-table}
\centering
\begin{tabular}{||c| c c||} 
\hline
Flux & in $x$ direction & in $y$ direction \\ [0.5ex] 
\hline
\hline
$x$-Momentum & $u^2 + p - \frac{1}{Re}\frac{\partial u}{\partial x}$ & $uv  - \frac{1}{Re}\frac{\partial u}{\partial y}$ \\
\hline
$y$-Momentum & $uv  - \frac{1}{Re}  \frac{\partial v}{\partial x}$ & $v^2 + p - \frac{1}{Re} \frac{\partial v}{\partial y}$\\
\hline
Mass & $u$ & $v$\\ 
[1ex]
\hline
\end{tabular}
\end{table}
Now, we explain the reason behind the communication time being higher in XPINN in comparison to cPINN.
For the present example of the incompressible Navier-Stokes equation, Table \ref{flux-table} gives the mass and momentum fluxes in $x$ and $y$ directions. 
From Table \ref{flux-table} it can be observed that the mass flux is represented by PDE variable $u, v$, which first saves one pass of computational graph for auto-differentiation and thus, reducing the communication time as shown in Figure \ref{cc}. However, the interface condition for XPINN requires the computation of residuals along the interfaces, which will require the explicit computation of the mass conservation equation, and subsequently performing the communication to pass the value in neighboring domains and therefore increasing the communication time justifying the results in Figure \ref{cc}. A similar explanation for the reduction in computation cost by the cPINN method can be given for the computation of interface conditions using momentum fluxes. 

So the main take away points from Figures \ref{cc}  and \ref{cc_gpu} are:
\begin{enumerate}
    \item Good weak scaling of the distributed cPINN and XPINN algorithms for the communication bound problem is achieved both with CPUs and GPUs.
    \item The communication time in XPINN is greater than cPINN in the case of spatial decomposition. The benefit of XPINN comes into play when spatial and temporal decompositions are performed simultaneously, as demonstrated in the example of Section \ref{st_vb}. 
\end{enumerate}
\begin{figure} 
\centering
\subfloat[]{\begin{tikzpicture} \label{cp_gpu}
\begin{axis}[
	scale=0.9,
	legend cell align=left,
	title={Runtime on Nodes, GPU Implementation},
	xlabel={Nodes},
	ylabel={Runtime for 100 iterations (sec)},
	xmin=2, xmax=26,
	ymin=0,ymax=4,
	ybar stacked,
	xtick={4,8, 12, 16, 20, 24},
	legend pos=north west,
	xmajorgrids=true,
	ymajorgrids=true,
	grid style=dashed,
] 
\addplot coordinates{(4, 1.3) (8, 1.32) (12, 1.3) (16, 1.22) (20, 1.35) (24, 1.22)};

\addplot+[
        error bars/.cd,
            y dir=both, 
            y explicit,
            error bar style={line width=1.0pt, color=black},
    ]coordinates{
         (4, 0.7)+-(0, 0.03) 
         (8, 0.75)+-(0, 0.0234)
         (12, 0.80)+-(0, 0.17)
         (16, 0.78)+-(0, 0.05)
         (20, 0.8)+-(0,0.03)
        (24, 0.78)+-(0,0.09)
};

\legend{Computation Time, Communication Time}
\end{axis}
\end{tikzpicture}} \ \ \ \ \ \ 
\subfloat[]{
\begin{tikzpicture}\label{xp_gpu}
\begin{axis}[
	scale=0.9,
	legend cell align=left,
	title={Runtime on Nodes, GPU Implementation},
	xlabel={Nodes},
	ylabel={},
	xmin=2, xmax=26,
	ymin=0,ymax=4,
	ybar stacked,
	xtick={4,8, 12, 16, 20, 24},
	legend pos=north west,
	xmajorgrids=true,
	ymajorgrids=true,
	grid style=dashed,
] 
\addplot coordinates{(4, 1.2) (8, 1.3) (12, 1.2) (16, 1.0) (20, 1.0) (24, 1.01)};
\addplot+[
        error bars/.cd,
            y dir=both, 
            y explicit,
            error bar style={line width=1.0pt, color=black},
    ]coordinates{
         (4, 1.3)+-(0, 0.0294) 
         (8, 1.4)+-(0, 0.0265) 
         (12, 1.8)+-(0, 0.029) 
         (16, 2.0)+-(0, 0.026) 
         (20, 2.0)+-(0,0.025) 
        (24, 2.0)+-(0,0.025) 
};
\legend{Computation Time, Communication Time}
\end{axis}
\end{tikzpicture}} 
\caption{Two-dimensional incompressible Navier-Stokes equations: Computation and communication time for (a) cPINN and (b) XPINN with 4000 residual points and 200 interface points.}
\label{cc_gpu}
\end{figure}
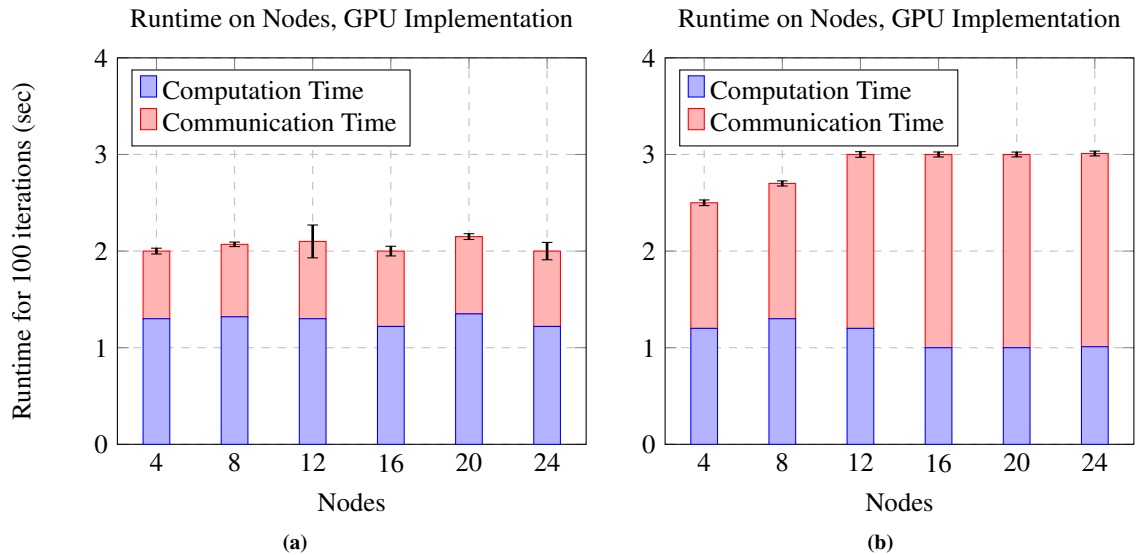
\begin{figure}
\begin{tikzpicture}\label{ws_cpinn}
\begin{axis}[
	xlabel = \# of GPUs,
	ylabel = Data points/sec,
	width=8cm,height=6cm,
    legend style={at={(0.01,0.87)}, anchor=west},
    xmin=2, xmax=28,
	ymin=0, ymax=2.5*10^5,
    ]
\addplot[color=red, mark=x, mark options={scale=1.25}, style=very thick] coordinates {
    (4, 34021.053)
    (8, 62361.0)
	(12, 94342.86)
    (16, 120581.82)
	(20, 151109.09)
	(24, 181636.36)
};
\addplot[color=blue, style= very thick, mark=+, mark options={scale=1.25}] coordinates {
     (4, 32601.00251)
     (8, 65202.00501)
	 (12, 97803.00752)
    (16,  130404.01)
	(20, 163005.0125)
	(24, 195606.015)
};
\legend{Observed: cPINN, Theoretical: cPINN}
\end{axis}
\end{tikzpicture}
\begin{tikzpicture}\label{ws_xpinn}
\begin{axis}[
	xlabel = \# of GPUs,
	ylabel = Data points/sec,
	width=8cm,height=6cm,
    legend style={at={(0.01,.87)}, anchor=west},
    xmin=2, xmax=28,
	ymin=0, ymax=1.5*10^5,
    ]
\addplot[color=red, mark=x, mark options={scale=1.25}, style=very thick] coordinates {
      (4, 21546.67)
     (8,  42245.16)
	 (12, 61912.50)
    (16, 85574.19)
	(20, 107238.71)
	(24, 137793.10)
    
};

\addplot[color=blue, style= very thick, mark=+, mark options={scale=1.25}] coordinates {
    (4, 22206.62959)
    (8, 44413.25918)
	(12, 66619.88877)
    (16, 88826.51835)
	(20, 111033.1479)
	(24, 133239.7775)
   
};
\legend{Observed: XPINN, Theoretical: XPINN}
\end{axis}
\end{tikzpicture} \ \ \ \ \ \ \ \ \ \ 
\caption{Two-dimensional incompressible Navier-Stokes equations: Weak GPU scaling for the distributed cPINN and XPINN algorithms.}
\label{weak_sclaing}
\end{figure}
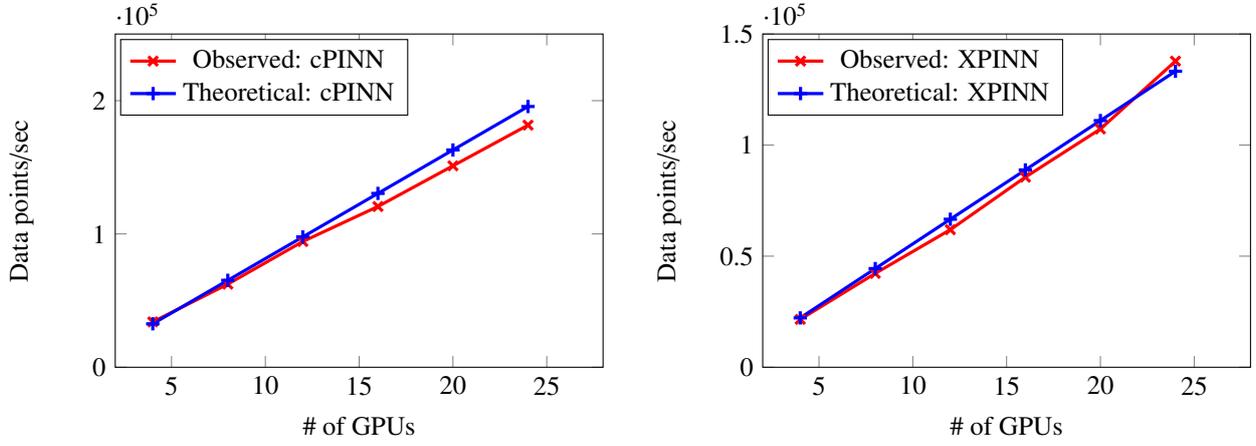

\subsubsection{Weak scaling and Strong scaling}
Next, we present the weak and strong scalings of distributed cPINN and XPINN algorithms. The weak scaling is performed on a heterogeneous architecture hosting CPUs and GPUs. Figures 8a and 8b represent the weak scaling for cPINN and XPINN, respectively. The weak scaling is measured by decomposing the domain from 4 to 24 subdomains with each subdomain sampled with 15000 residual points, 1000 interface points, and 80 boundary points, and the number of data points increases as the number of GPUs increases to balance the load equally. The hardware architecture is the same as that used in Figure \ref{cc_gpu}. The weak scaling is computed by using the equation $\ref{wse}$. Figures 8a and 8b represent a linear relation between the number of data points processed per second and the number of nodes, which also conforms accurately with the theoretical performance. From these figures, we note that the number of data points processed per second is more for cPINN than XPINN due to the larger communication time in XPINN. 
\begin{figure}
\centering
\subfloat[] {\begin{tikzpicture}\label{ss_su}
\begin{axis}[
	scale=0.9,
	legend cell align=left,
	xlabel={Nodes},
	ylabel={Speed up},
	xmin=2, xmax=26,
	ymin=0,ymax=22,
	ybar,
	ybar,
    ymin=0,
	xtick={4,8, 12, 16, 20, 24},
	legend pos=north west,
	xmajorgrids=true,
	ymajorgrids=true,
	grid style=dashed,
] 
\addplot coordinates{(4, 4) (8, 6.04) (12, 8.40) (16, 12.42) (20, 15.87) (24, 20) };
\addplot coordinates{(4, 4) (8, 6.30) (12, 7.30) (16, 9.49) (20, 15.02) (24, 16.70) };
\legend{cPINN, XPINN}
\end{axis}
\end{tikzpicture}} \ \ \ \ \ \ \ \ \ \ 
\subfloat[]{
\begin{tikzpicture}\label{eff}
\begin{axis}[
	scale=0.9,
	legend cell align=left,
	xlabel={Nodes},
	ylabel={Efficiency},
	xmin=2, xmax=26,
	ymin=0,ymax=1.4,
	ybar ,
	xtick={4,8, 12, 16, 20, 24},
	legend pos=north west,
	xmajorgrids=true,
	ymajorgrids=true,
	grid style=dashed,
] 
\addplot coordinates{(4, 1) (8, 0.76) (12, 0.70) (16, 0.78) (20, 0.7935) (24, 0.83) };
\addplot coordinates{(4, 1) (8, 0.79) (12, 0.61) (16, 0.60) (20, 0.751) (24, 0.70)};
\draw[dashed, ultra thick] (axis cs:\pgfkeysvalueof{/pgfplots/xmin},1) -- (axis cs:\pgfkeysvalueof{/pgfplots/xmax},1);
\legend{cPINN, XPINN}
\end{axis}
\end{tikzpicture}} 
\caption{Two-dimensional incompressible Navier-Stokes equations: Strong scaling bar plots where the left figure gives speedup and the right figure gives efficiency, respectively for both cPINN and XPINN methods.}
\label{ss_cxpinn}
\end{figure}
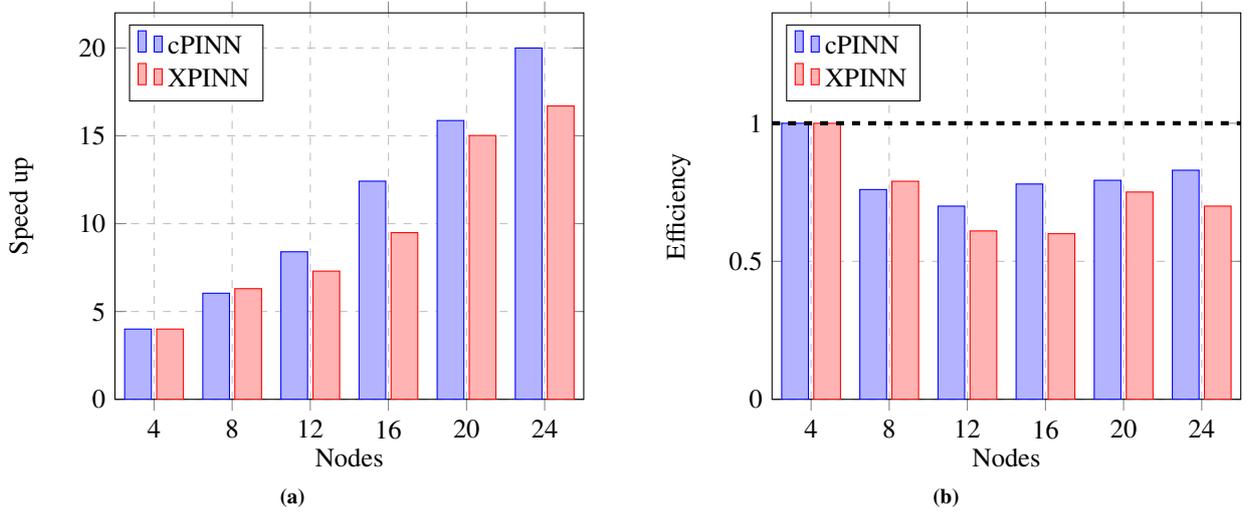

Now, we present the strong scaling for cPINN and XPINN in Figures \ref{ss_cxpinn} measured on \textit{Intel's Cascade lake CPUs}. The performance shown in Figure \ref{ss_su} is measured by fixing the problem size with total data points 249600, which includes boundary, residual, and interface data points. Similar to the earlier examples, each rank is mapped node-wise. Figure \ref{ss_su} represents the speedup for cPINN and XPINN, which shows a linear relationship as the number of nodes increases, which eventually reduces the work per processor. Figure \ref{ss_su} also describes that the speedup of XPINN is less than cPINN due to more communication in XPINN.  Figure \ref{eff} shows the efficiency of cPINN and XPINN algorithms computed using equation \ref{sse}. The results show that cPINN is more efficient with an average efficiency of 80\%, while XPINN achieves only 70\% efficiency. The issue of enhancing the efficiency of the XPINN method is addressed in Section\ref{st_vb}.

 \subsubsection{Comparison between the runtime obtained from data-parallel vanilla PINN, cPINN and XPINN}
We also present  a comparison between runtimes obtained from vanilla PINN method using the data-parallel approach, and  parallel cPINN as well as XPINN methods to solve equation (\ref{NSEq}). In Table \ref{CompVCX}, we show the runtime for 200 iterations. The hardware architecture and software stack is the same as those used in Figure \ref{cc_gpu}. To perform the test, we used uniform network architecture for all the three cases with 8 hidden-layers and 64 neurons in each layer.

\begin{table}
 
\caption{Comparison of walltimes averaged over 10 runs (mean and standard deviation) for first 200 iterations using the data-parallel vanilla PINN, cPINN and XPINN methods.}
\label{CompVCX}
\centering

\begin{tabular}{||c| c | c | c||} 
\hline
$\#$ GPU's (No. of Nodes) & Data-parallel PINN & cPINN & XPINN \\ [0.5ex] 
\hline
\hline
4 (1) & 14.49 $\pm$ 0.34 & 18.76 $\pm$ 0.12 & 19.77 $\pm$ 0.34 \\
\hline
8 (2) & 15.16 $\pm$ 0.56 & 19.22 $\pm$ 0.20 & 20.73 $\pm$ 0.15 \\ 
\hline
12 (3) & 16.93 $\pm$ 0.74 & 19.48 $\pm$ 0.07 & 20.97 $\pm$ 0.15 \\
\hline
16 (4) & 17.63 $\pm$ 0.80 &  19.71	$\pm$ 0.07 & 21.34 $\pm$ 0.05 \\ 
\hline
20 (5) & 18.00 $\pm$ 0.49 & 19.84 $\pm$ 0.02 & 21.48 $\pm$ 0.05\\ 
\hline
24 (6) & 18.03 $\pm$ 0.90 & 19.74 $\pm$ 0.08
& 21.63 $\pm$ 0.07  \\ 
\hline
\end{tabular}
\end{table}

To have a fair comparison between data-parallel PINN, cPINN and XPINN methods, we saturated the GPUs by using large number of residual points resulting into $\approx$ 98\% volatile utilization of GPUs. These runtimes are measured in a weak scaling sense by decomposing the domain from 4 to 24 subdomains with each subdomain endowed with 50000 residual points, 500 interface points, and 80 boundary points, and the number of data points increases as the number of GPUs increases to balance the load equally. The runtimes in Table \ref{CompVCX} is for the computations carried out with Float32 and each node consists 4 GPUs. To remove biases in runtimes, we present walltimes averaged over 10 runs for data-parallel PINN, cPINN and XPINN along with their standard deviations. Table \ref{CompVCX} shows that the runtimes for data-parallel PINN and cPINN are almost similar despite the fact that cPINN requires computations of fluxes. 
We also note that the data-parallel requires communication through \textit{allreduce} collectives which is expensive and has incremental overhead as number of processors (GPUs) increases, as reflected in Table \ref{CompVCX}. However, in cPINN and XPINN the communication time will be fixed with respect to number of processors (GPUs) as these methods will require point-to-point communication and will have 4- and 6-way communication in 2D and 3D spaces (Cartesian grid), respectively. The runtimes for XPINN are slightly higher than cPINN but this can be significantly reduced by partitioning the domain in temporal dimension as well, which is already deliberated in Section \ref{st_vb}. Moreover, as discussed earlier, data-parallel has only single network to train as opposed to multiple networks used in cPINN and XPINN methods, which are additionally endowed with local representation capacity. Furthermore, data-parallel approach is not adequate for the problems involving multi-physics data for the real world problems with heterogeneous physics.

\subsection{Viscous Burgers equation with space-time domain decomposition}
\label{st_vb}
This test case aims to show the advantage of XPINN over cPINN by decomposing the domain along the time axis. To this end, we again consider a one-dimensional viscous Burgers equation given by
\begin{equation}\label{BE1}
u_t + u u_x - \nu u_{xx}=0, ~~~ x\in [-1, 1],~t>0,
\end{equation}
with initial and boundary conditions $u(0, x) = -\sin(\pi x)$ and  $u(t, -1) = u(t,1)=0$, respectively, and $\nu = 0.01/\pi$.
  \vspace*{0.5cm}
  
  \begin{minipage}{\textwidth}
  \begin{minipage}[b]{0.4\textwidth}
   \centering
\includegraphics[trim=0cm 0cm 0cm 0cm, clip=true, scale=0.375, angle = 0]{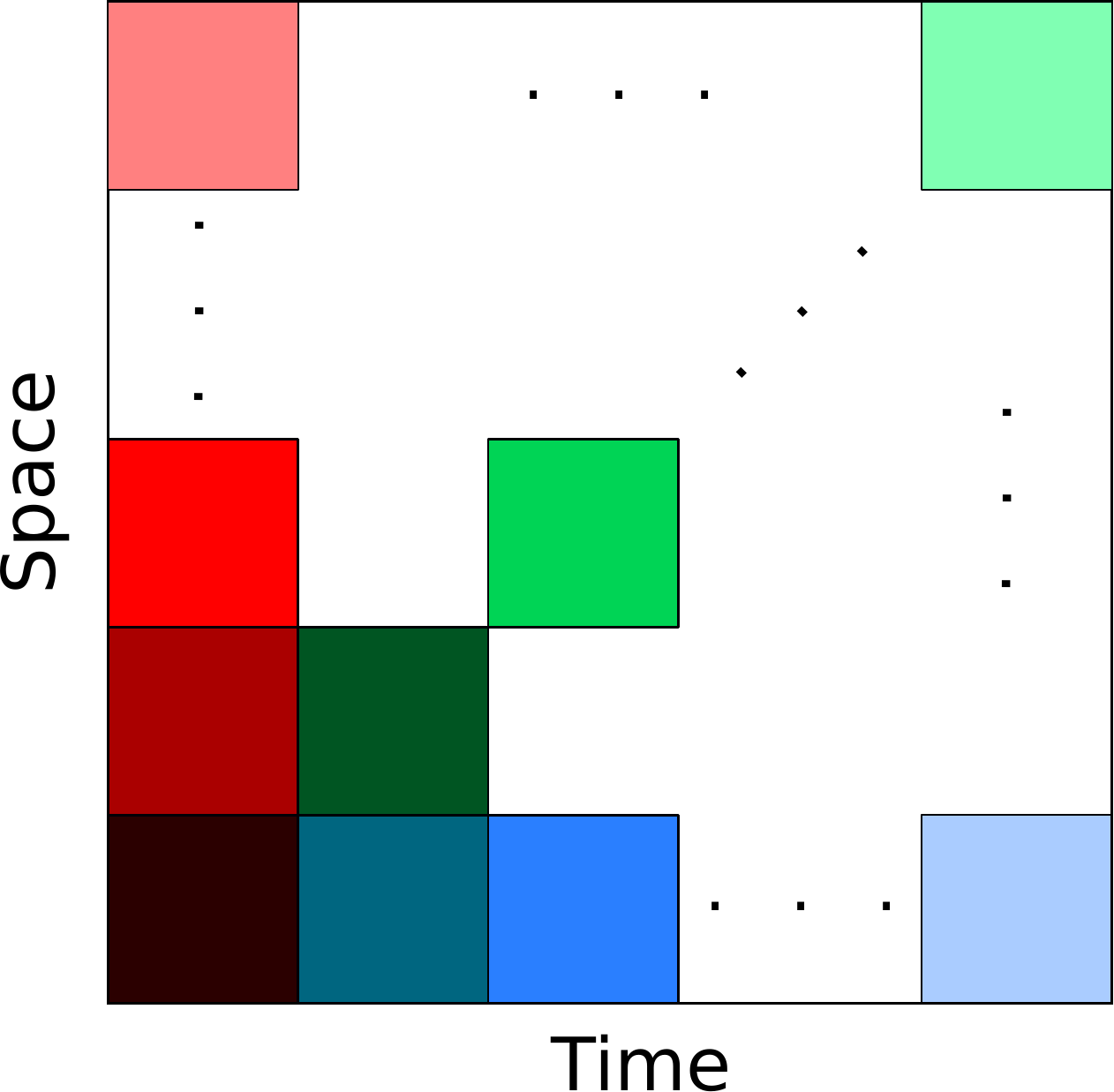}
    \captionof{figure}{Viscous Burgers equation: Schematic representation of space-time decomposition in XPINN methodology.}
    \label{fig:VB}
  \end{minipage}
  \hfill
  \begin{minipage}[b]{0.55\textwidth}
    \centering
\begin{tabular}{ |c|c|c|c|  }
 \hline
 \multicolumn{4}{|c|}{cPINN vs XPINN partition} \\
 \hline
$\#~ x-$  & $\#~ t$  & cPINN time& XPINN time \\
 partitions &  partitions & per iter. (s) & per iter. (s)\\
 \hline
4   &  1    & 0.14 & - \\
8   &  1    & 0.12 & - \\
16   &  1    & 0.12 & - \\
32   &  1    & 0.125 & - \\
2   &  2   & - &  0.064 \\
4   &  2   & - &  0.060 \\
4   &  4   & - &  0.060 \\
8   &  4   & - &  0.060 \\
 \hline
\end{tabular}
      \captionof{table}{Viscous Burgers equation: Comparison of cPINN and XPINN average wall time per iteration for different number of spatio-temporal domain divisions.}
      \label{TableVB}
    \end{minipage}
  \end{minipage}
  \vspace*{0.5cm}
  
In each subdomain, we used a neural network with 5 hidden-layers and 20 neurons per layer. In each network, the activation function is hyperbolic tangent and the learning rate is 8e-4. The total number of residual points is 80k in the whole domain, which is fixed for all cases. Also, the number of interface points is 20 along time as well as space directions. As the domain is subdivided into subdomains, the residual points, as well as interface points are also distributed uniformly in each subdomain. In the case of cPINN, the domain is divided only along the spatial axis whereas, in the XPINN, space-time domain decomposition can be performed as shown schematically in figure \ref{fig:VB}. Table \ref{TableVB} shows the weak scaling on CPUs for both cPINN and XPINN. In the case of cPINN, the domain is partitioned along the space into 4, 8, 16, and 32 subdomains whereas in the case of XPINN, both space and time axes are partitioned in such a way that the total number of subdomains remains the same in both cases. The table shows the average computational time for cPINN and XPINN per iteration, which is higher for cPINN. This is because in the cPINN implementation, the buffer size for communication is static along the time direction and results in more communication overhead. Moreover, in cPINN, the computation of interface loss requires gradient computation for all the interface points, which further increases computational cost. However, in the XPINN implementation, the buffer size for communication is divided by the number of subdomains created along the space-time axes. 

This example shows the efficacy of the XPINN method when the domain decomposition is performed along time direction. Thus, XPINN is a very promising deep learning approach for transient problems.

\subsection{Inverse problem: Steady-state heat conduction with variable conductivity}
The governing equation for the steady-state heat conduction problem is written as
\begin{equation}\label{heq}
 \partial_x (K(x,y)T_x) + \partial_y(K(x,y) T_y) = f(x,y)
\end{equation}
with Dirichlet boundary conditions for temperature $T$ and thermal conductivity $K$, obtained from the exact solution. The forcing term $f(x,y)$ is obtained from the exact solution for variable thermal conductivity and temperature, which is assumed to be of the form
\begin{align*}
 T(x,y) &= 20~\text{exp}(-0.1y) 
 \\ K(x,y) &= 20 + \text{exp}(0.1y)~\sin(0.5x).
\end{align*}

\begin{figure} [htpb] 
\centering
\includegraphics[trim=0.5cm 0cm 0cm 0cm, clip=true, scale=0.24, angle = 0]{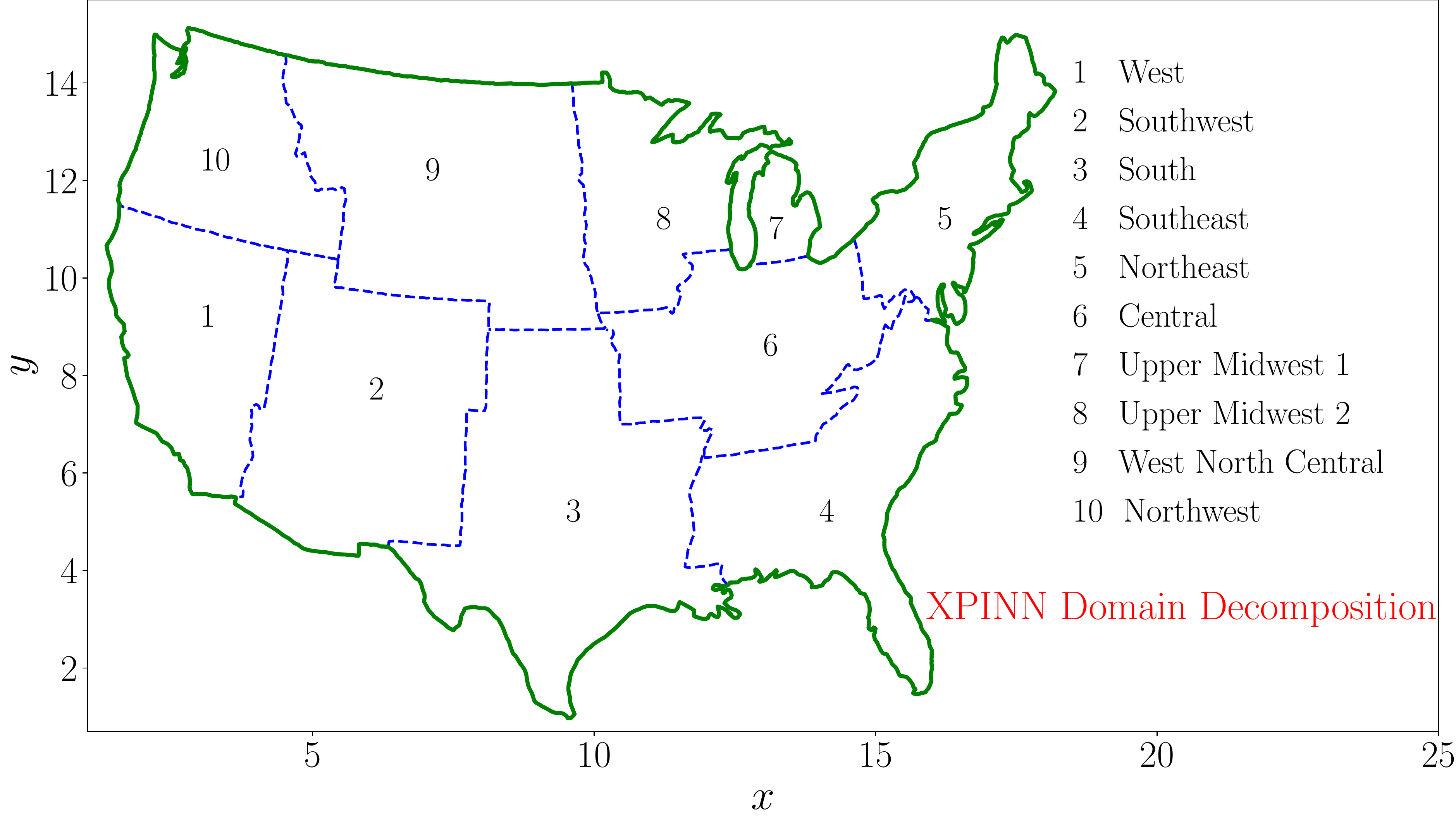}
\includegraphics[trim=0cm 0cm 0cm 0cm, clip=true, scale=0.313, angle = 0]{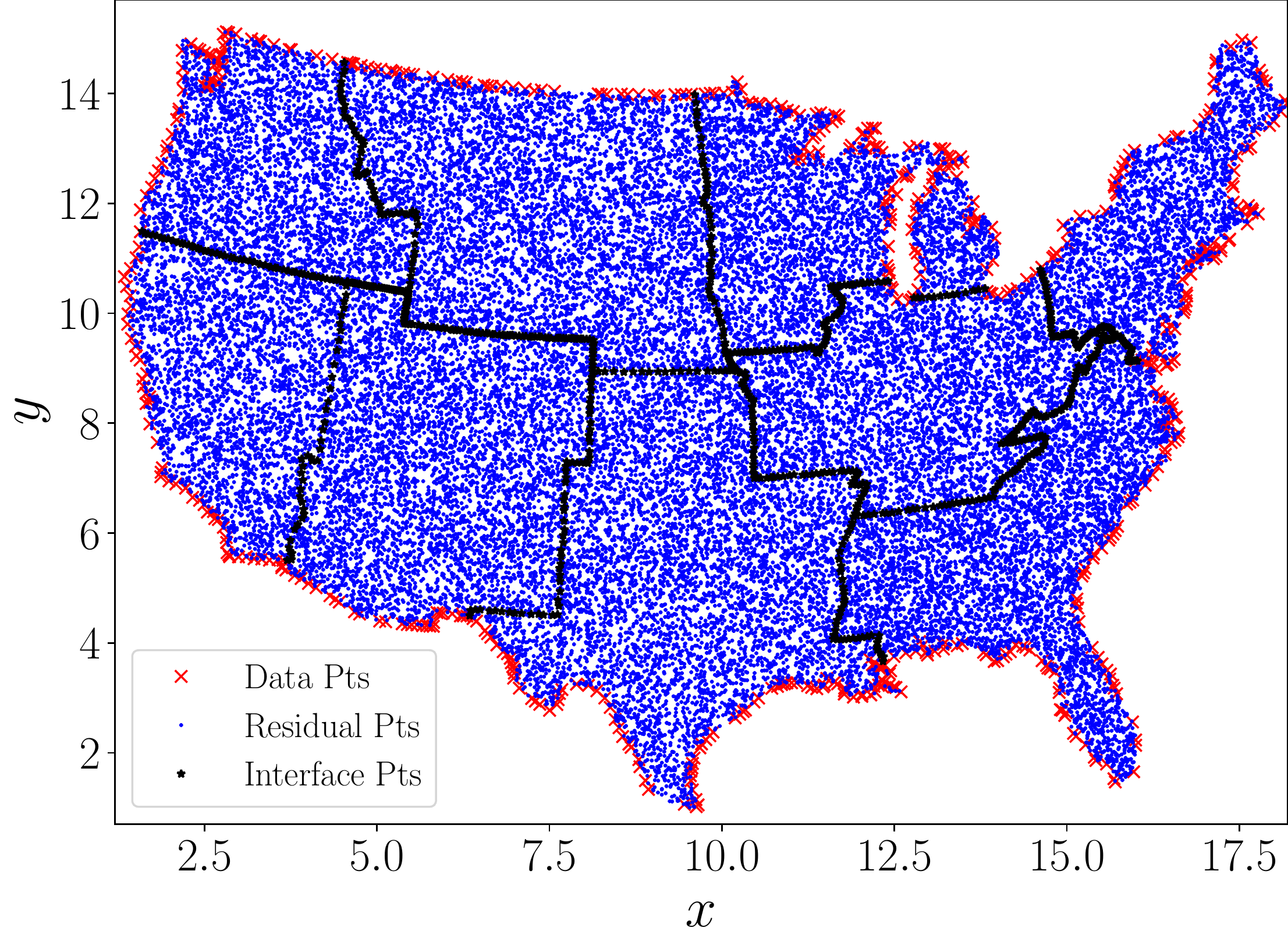}
\caption{Steady-state heat conduction with variable conductivity: Domain decomposition of the US map into 10 regions (left) and the corresponding data, residual, and interface points in these regions (right). }
\label{fig:USmapD}
\end{figure}

Materials whose thermal conductivity is a function of space are the so-called \textit{functionally graded materials}, which are a particular type of composite materials. The material domain is chosen to be a map of the United States divided into 10 subdomains as shown in figure \ref{fig:USmapD}. The interfaces between these subdomains are shown by the dashed blue line whereas the domain boundary is shown by the solid green line. The boundary and the training data is available for every subdomain. Unlike cPINN, XPINN can easily handle such complex subdomains with arbitrarily shaped interfaces, thus, providing greater flexibility in terms of domain subdivision. Figure \ref{fig:USmapD}(right) shows the residual, training data, and interface points in blue dot, red cross, and black asterisk, respectively, over the entire domain.
\begin{figure} [htpb] 
\centering
\includegraphics[trim=0cm 0cm 0cm 0cm, clip=true, scale=0.25, angle = 0]{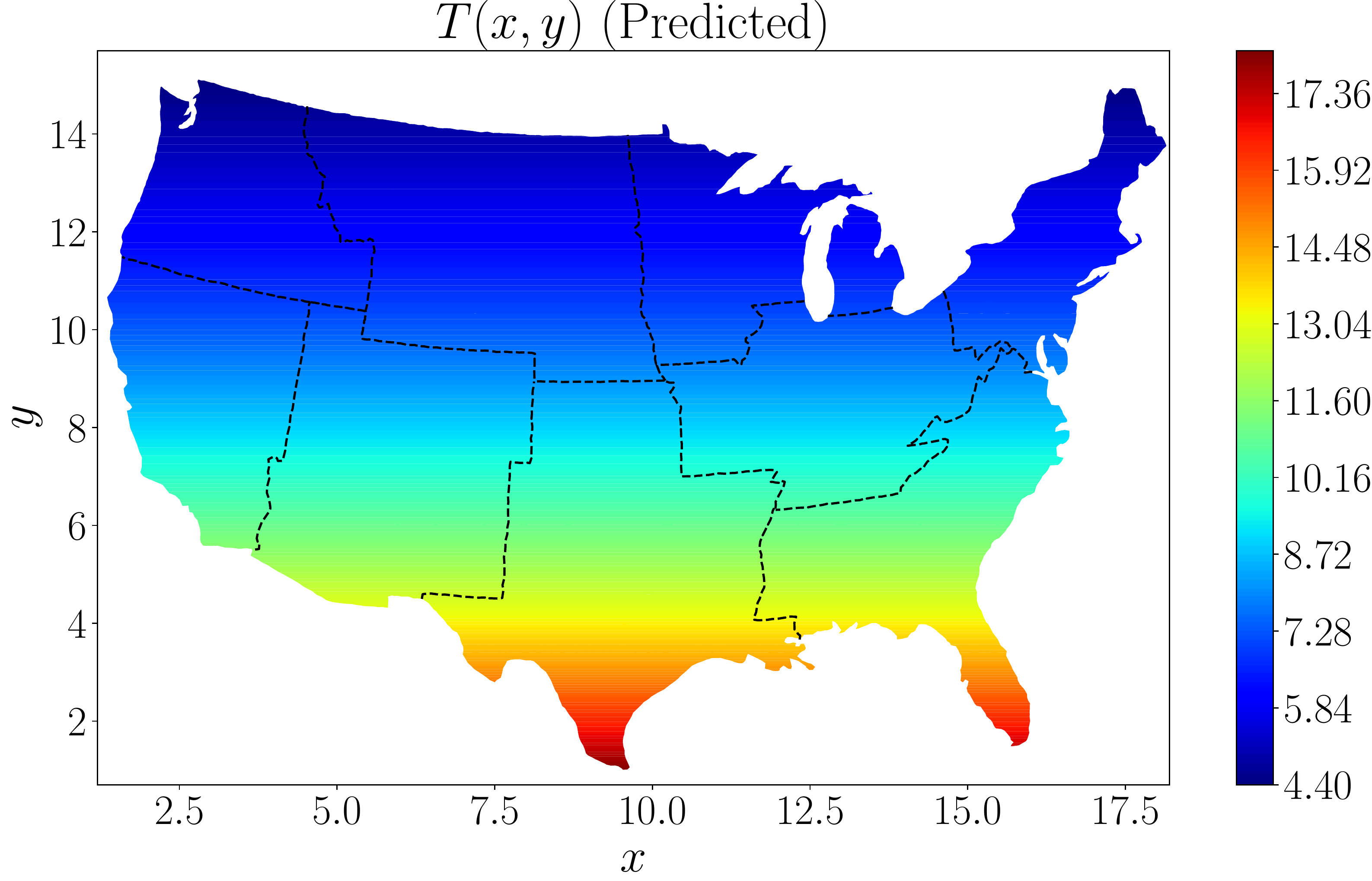}
\includegraphics[trim=0cm 0cm 0cm 0cm, clip=true, scale=0.25, angle = 0]{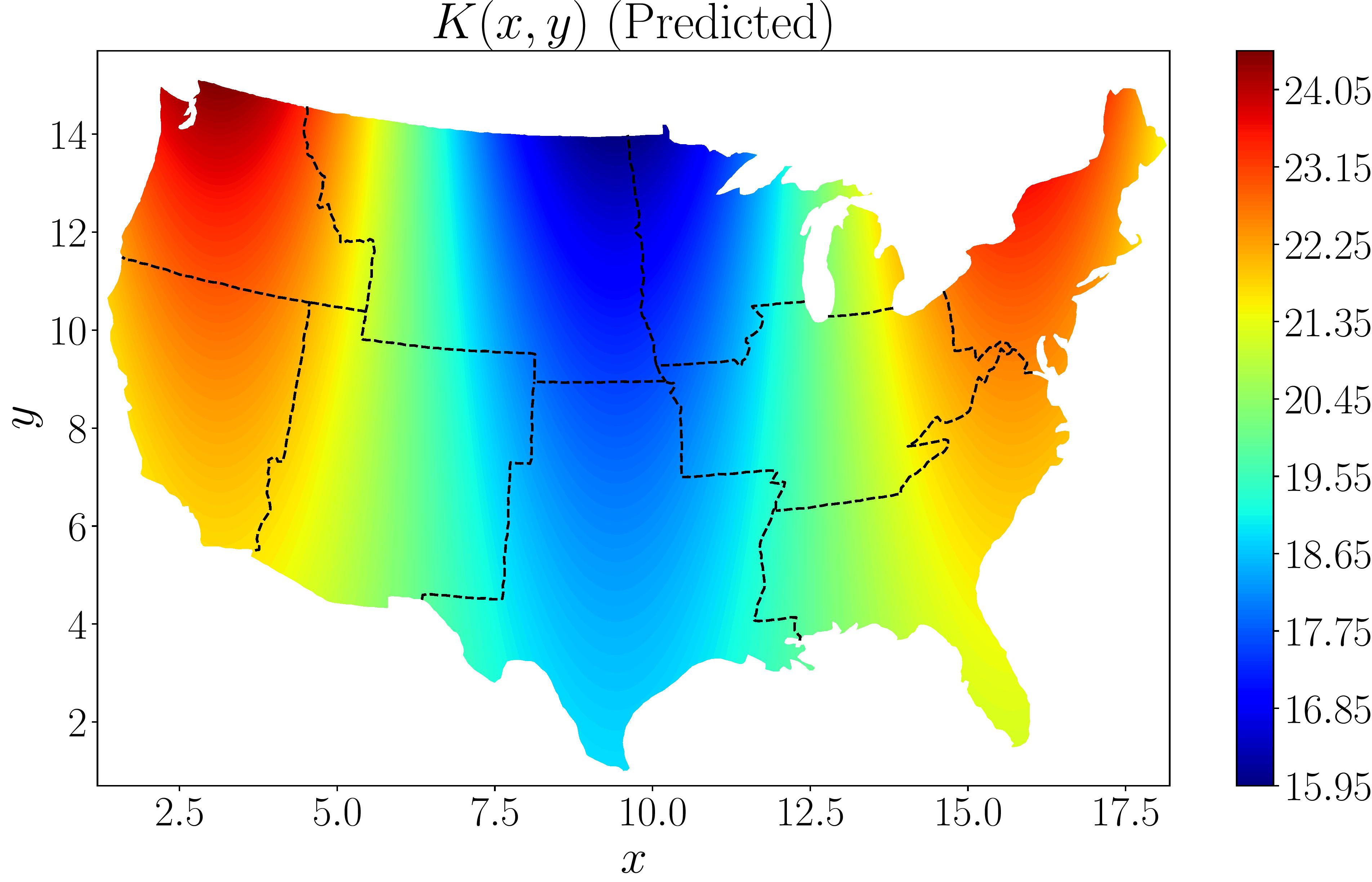}

\includegraphics[trim=0cm 0cm 0cm 0cm, clip=true, scale=0.088, angle = 0]{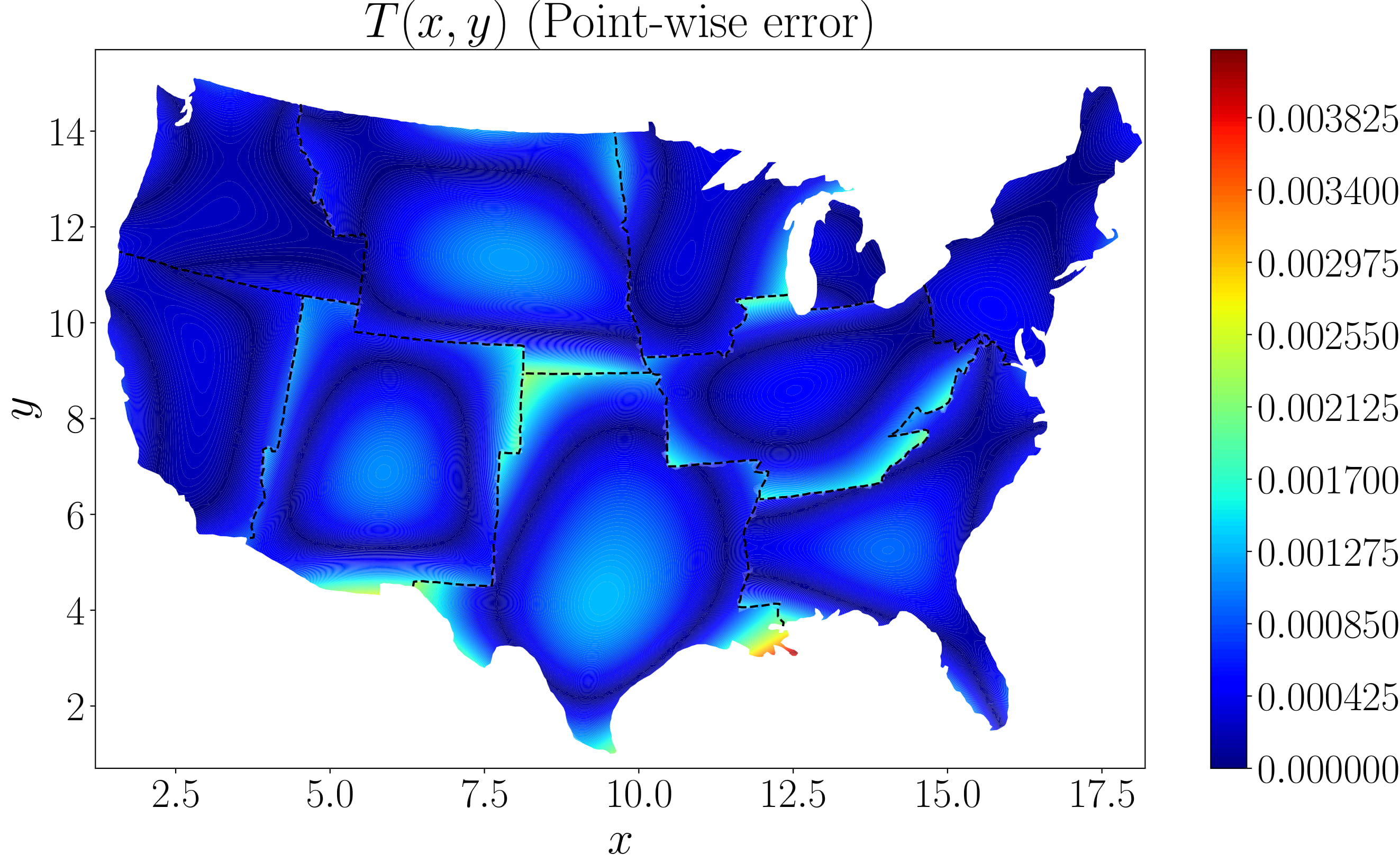}
\includegraphics[trim=0cm 0cm 0cm 0cm, clip=true, scale=0.088, angle = 0]{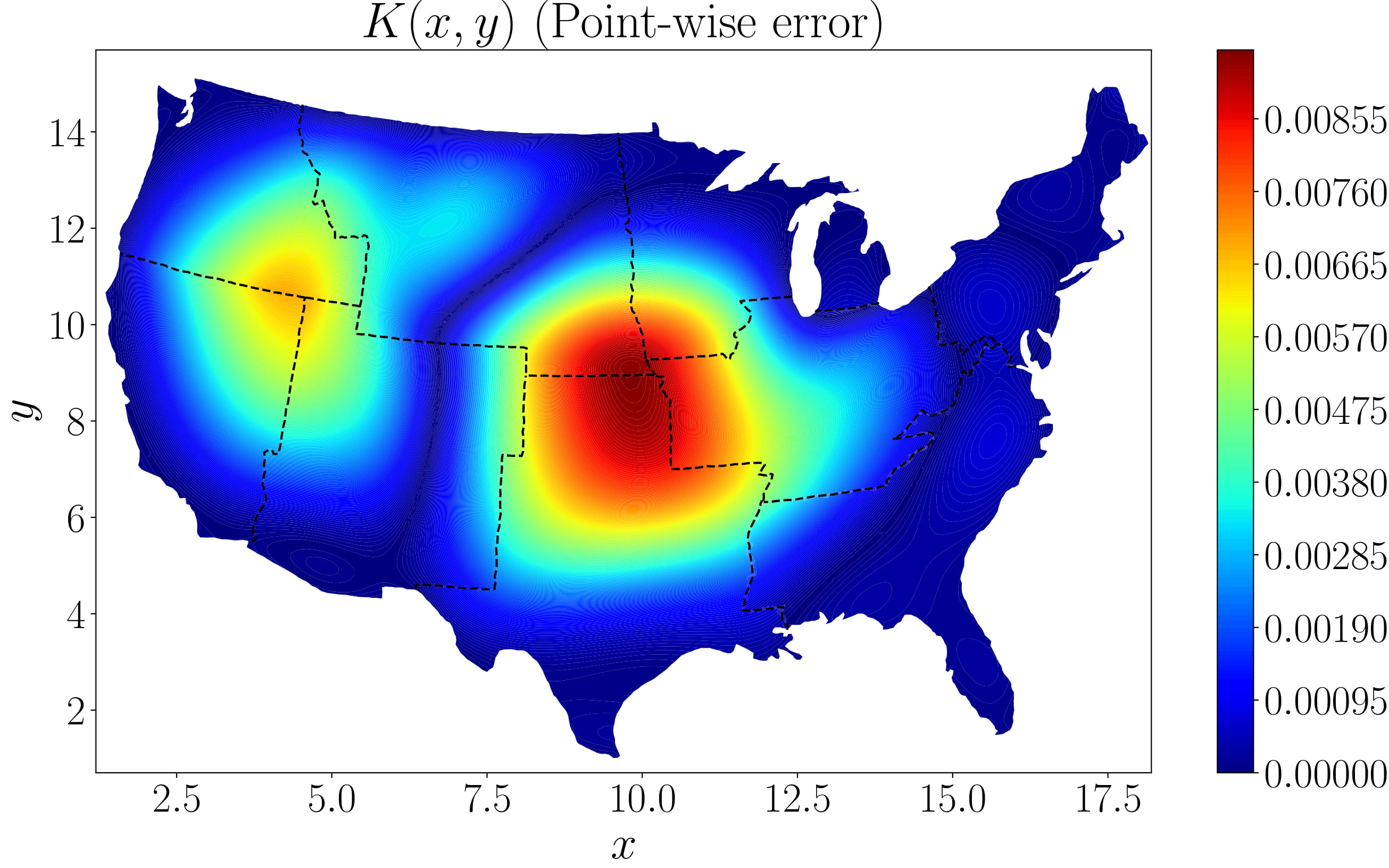}
\caption{Steady-state heat conduction with variable conductivity: The first row shows the contour plots for the predicted temperature $T(x,y)$ and thermal conductivity $K(x,y)$ while the second row shows the corresponding absolute point-wise errors.}
\label{fig:USmap_Sol}
\end{figure}

In this inverse problem, the temperature is assumed to be known in the domain but the variable thermal conductivity is unknown. The aim is to infer the unknown thermal conductivity of the material from a few data points for $K$ available along the boundary line and temperature values available inside the domain.
We employed a single PINN in each subdomain, and the details about the network's architecture are given in the table \ref{Table_USmap}.
\begin{table}[htpb]
\begin{center}
\small \begin{tabular}{ccccccccccc} \hline 
 Subdomain number &  1 & 2&  3 & 4&  5 & 6&  7 & 8&  9 & 10
 \\ \hline
$\#$ Residual points & 3000 &4000& 5000 &4000& 3000 &4000& 800 &3000& 5000 &4000
 \\
 Adaptive Activation function & $\tanh$ &$\sin$ & $\cos$ & $\tanh$ &$\sin$ & $\cos$ & $\tanh$ &$\sin$ & $\cos$ & $\tanh$
 \\
 \hline 
 \end{tabular}
\caption{Steady-state heat conduction with variable conductivity: Neural network architecture in each subdomain.}\label{Table_USmap}
\end{center}
\end{table}
In each subdomain, we used 3 hidden-layers with 80 neurons in each layer, and the learning rate is 6e-3, which is fixed for all networks. 

Figure \ref{fig:USmap_Sol} (top row) shows the predicted values of temperature and thermal conductivity. The absolute point-wise errors are given in the bottom row. 
The XPINN method accurately inferred the thermal conductivity, which shows that XPINN can easily handle highly irregular and non-convex interfaces.
\begin{figure}
\centering
\subfloat[]{
\begin{tikzpicture}\label{heat_cpu}
\begin{axis}[
	scale=0.8,
	legend cell align=left,
	xlabel={\# Nodes-CPUs},
	ylabel={Walltime (hours)},
	ymin=0,ymax=15,
	ybar,
	ybar,
    ymin=0,
    xtick=data,
    enlarge x limits={abs=1.5cm},
    symbolic x coords={1-CPU, 10-CPUs},    
	legend pos=north east,
	xmajorgrids=true,
	ymajorgrids=true,
	grid style=dashed,
] 
\addplot coordinates{(1-CPU, 4.60) (10-CPUs, 0.5)};
\node[draw=none, color=blue] at (-7, 52.5) {1x};
\node[draw=none, color=blue] at (92.5, 11.2) {9x};
\addplot coordinates{(1-CPU, 12.83) (10-CPUs, 1.33)}; 
\node[draw=none, color=red] at (5, 134.0) {1x};
\node[draw=none, color=red] at (105, 20.5) {10x};
\legend{Float32, Float64}
\end{axis}
\end{tikzpicture}
} \ \ \ \ \ \ \ \ 
\subfloat[]{
\begin{tikzpicture}\label{heat_gpu}
\begin{axis}[
	scale=0.8,
	legend cell align=left,
	xlabel={\# Nodes-GPUs},
	ylabel={},
	ymin=0,ymax=6,
	ybar,
	ybar,
    ymin=0,
    xtick=data,
    enlarge x limits={abs=1.5cm},
    symbolic x coords={1-GPU, 10-GPUs},    
	legend pos=north east,
	xmajorgrids=true,
	ymajorgrids=true,
	grid style=dashed,
] 
\addplot coordinates{(1-GPU, 3.3) (10-GPUs, 0.46)};
\node[draw=none, color=blue] at (-8, 360.5) {1x};
\node[draw=none, color=blue] at (93,70.5) {7x};
\addplot coordinates{(1-GPU, 4.16) (10-GPUs, 0.46)};
\node[draw=none, color=red] at (8, 450.5) {1x};
\node[draw=none, color=red] at (106,70.5) {9x};
\legend{Float32, Float64}
\end{axis}
\end{tikzpicture}
}
\caption{Steady-state heat conduction with variable conductivity: Walltime and speedup of parallel XPINN algorithm on CPUs and GPUs implemented for the inverse heat conduction problem in (\ref{heq}); (a) speedup and wall time for the parallel XPINN code on Intel's Xeon(R) Gold 6126 CPU. The speed and wall time is measured for computations performed with single (Float32) and double-precision numbers (Float64); (b) speedup and wall time, measured for single- and double-precision operations, on Nvidia's V100 GPUs.}
\label{xpinn_heat_sl}
\end{figure}
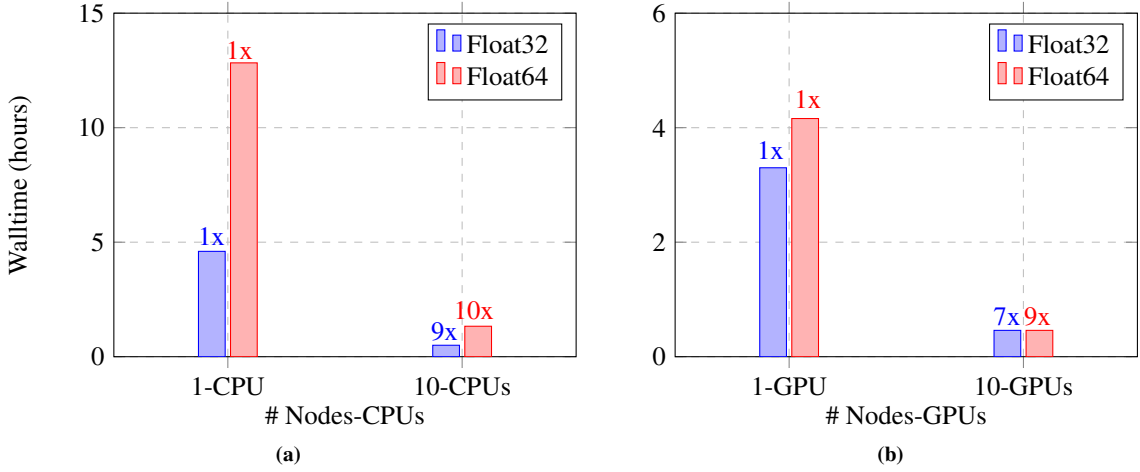
Figure \ref{xpinn_heat_sl} represents the performance of the parallel XPINN algorithm deployed for solving the inverse heat conduction problem. The performance is measured for the data points given in Table \ref{Table_USmap}. Figure \ref{heat_cpu} represents the wall time and scaling of the parallel XPINN algorithm on CPUs by computing the solution in each domain using one CPUs. The continuity of solution on the shared boundaries of the subdomain is imposed by passing the solution on each domain boundary to the neighboring subdomains though MPI protocols. Here, each CPU corresponds to a rank mapped by node, and Intel's Xeon(R) Gold 6126 CPUs are used for measurement. First, we measured the wall time of the algorithm on one CPU and considered it as baseline data $(1\text{X})$ for the scaling. Thereafter, we computed the wall time for 10 CPUs, leading to a $(9\text{X})$ and  $(10\text{X})$ for single (32-bit float) and double-precision (64-bit float) computation, respectively. The scaling for double-precision is relatively better as the communication process is shadowed by the computation time due to more time being spent on double-precision arithmetic. However, double-precision-based computation increases wall time by a factor of $2.5$ for single and multiple CPU-based implementations.

Next, we report the walltime and GPU-based implementation. The algorithm is implemented on Nvidia V100 GPUs with 16 GB memories. The hardware architecture is similar to that reported in the example of incompressible Navier-Stokes equations. Figure {\ref{heat_gpu}} represents the walltime for one and ten GPUs  for single- and double-precision arithmetic. On a single GPU, considered as $(1\text{X})$ model the wall time for double-precision arithmetic 30\% more than that received for single-precision. In the multi-GPU implementation, once GPU is used for each subdomain and here one rank mapped by the node corresponds to the combination of 1CPU + 1GPU. The walltime for 10-GPUs (10 Nodes or 10 ranks) yields a scaling $(7\text{X})$ and $(9\text{X})$ for single- and double-precision arithmetic. We note that the residual points in each subdomain are not enough to saturate the GPUs and therefore more time is spent on fetching the data to memory and inter-GPU communication. To provide more intuition in the purview of the current implementation, for a typical V100 (16 GB) memory, the single-precision performance is 14 TFLOPs with a memory bandwidth of 900 GB/s and therefore for each byte of transfer (memory to GPU core) 15.6 instruction (FLOPs/ Bandwidth) needs to be issued to occupy the GPUs completely. In the context of the current problem, the partition and therefore the load on GPUs (or CPUs) are static and does not change throughout the computation and thus, subdomain 7, endowed with only 800 residual points, has to wait until all the GPUs (or CPUs) complete their work for the respective subdomain. This also results in slight sub-linear performance. However, the performance presented in Figure \ref{xpinn_heat_sl} shows a very good linear scaling for CPUs on a heterogeneous (CPU + GPU) architecture. Additionally, the scaling of the presented algorithm on CPUs only architecture concurs with the idea of Daghaghi \etal \cite{daghaghi2021accelerating}, where the authors try to revisit the algorithms used in machine learning to make them faster on CPUs rather than getting fixated with one-dimensional development of specific hardware to run the matrix multiplication faster.    

 In this test case, the partition of the domain is performed  by manually choosing the interface points. This was done to show the efficacy of the XPINN algorithm for non-convex or irregular subdomains. In such complex subdomains, the distribution system often faces the problem of load imbalance, which can seriously degrade the performance of the system. A more efficient approach could be utilized to decompose the domain such that the subdomains are packed optimally for inter-node or inter-process communication. A suitable point cloud \cite{Rusu_ICRA2011_PCL}  or K-way partition \cite{METIS} based approach will result in further optimizing the communication.

\subsection{A note on the choice of cPINNs and XPINNs methods for practical problems}
Both cPINN and XPINN have distinct advantages over each other. From the computational experiments, it can be observed that cPINN has lower communication overhead than XPINN but it can only be applied to physical problems like conservation laws. On the other hand, unlike cPINN, XPINN can be used for spatio-temporal decomposition with universal applicability irrespective of the nature of physical laws. In practice, both can be employed simultaneously, for example, in case of time-dependent conservation laws, cPINN can be employed in space, whereas XPINN can be used to decompose in time direction. Such a combination can drastically reduce the computational power associated with the training of the neural networks for scientific problems.  Moreover, such a combined approach efficiently balances the communication overhead for all the networks.

\section{Summary}
We have developed a parallel framework for the domain decomposition-based conservative and extended physics-informed neural networks abbreviated as cPINNs and XPINNs, respectively. Here, we have presented a hybrid parallel algorithm for cPINN and XPINN constructed with a programming model described by MPI $+$ X, where X $\in \{\text{CPUs},~\text{GPUs}\}$. The main advantage of the parallel cPINN and XPINN over more classical data and model parallel approaches is their adaptivity for all hyperparameters of the neural network in each subdomain. We compared the performance of distributed cPINN and XPINN methodologies for both forward and inverse problems, presenting both weak scaling and strong scaling results. We have also shown the speedup and efficiency of the algorithm for both cPINN and XPINN methods. In all the computational experiments we observed that the spatial decomposition contributes to greater communication overhead for the XPINN in comparison to the cPINN, but we also showed that the efficiency of the XPINN method can be increased by decomposing the domain along the time axis, which cannot be done in the case of cPINN. Moreover, in terms of implementation of interface conditions, XPINN can be easily employed in more complex subdomains as shown by the inverse heat conduction problem over the US map.

\section*{Acknowledgement}
The work was supported by OSD/AFOSR MURI Grant FA9550-20-1-0358, the DOE PhILMs project (DE-SC0019453) and the DARPA CompMods grant HR00112090062.
The weak scaling and GPUs based simulations were primarily performed at Oak Ridge
Leadership Computing Facility (OLCF) through the OLCF Director’s
Discretion Program under project ENG120, which is supported
by the Office of Science of the U.S. Department of Energy under
Contract DE-AC05-00OR22725. This research was partially conducted using computational resources and services at the Center for Computation and Visualization (OSCAR), Brown University.

\bibliography{ref}

\begin{thebibliography}{10}
\expandafter\ifx\csname url\endcsname\relax
  \def\url#1{\texttt{#1}}\fi
\expandafter\ifx\csname urlprefix\endcsname\relax\def\urlprefix{URL }\fi
\expandafter\ifx\csname href\endcsname\relax
  \def\href#1#2{#2} \def\path#1{#1}\fi

\bibitem{bojarski2016end}
M.~Bojarski, D.~Del~Testa, D.~Dworakowski, B.~Firner, B.~Flepp, P.~Goyal, L.~D.
  Jackel, M.~Monfort, U.~Muller, J.~Zhang, et~al., End to end learning for
  self-driving cars, arXiv preprint arXiv:1604.07316.

\bibitem{hinton2012deep}
G.~Hinton, L.~Deng, D.~Yu, G.~E. Dahl, A.-r. Mohamed, N.~Jaitly, A.~Senior,
  V.~Vanhoucke, P.~Nguyen, T.~N. Sainath, et~al., Deep neural networks for
  acoustic modeling in speech recognition: The shared views of four research
  groups, IEEE Signal processing magazine 29~(6) (2012) 82--97.

\bibitem{huang2014historical}
X.~Huang, J.~Baker, R.~Reddy, A historical perspective of speech recognition,
  Communications of the ACM 57~(1) (2014) 94--103.

\bibitem{litjens2017survey}
G.~Litjens, T.~Kooi, B.~E. Bejnordi, A.~A.~A. Setio, F.~Ciompi, M.~Ghafoorian,
  J.~A. Van Der~Laak, B.~Van~Ginneken, C.~I. S{\'a}nchez, A survey on deep
  learning in medical image analysis, Medical image analysis 42 (2017) 60--88.

\bibitem{khandani2010consumer}
A.~E. Khandani, A.~J. Kim, A.~W. Lo, Consumer credit-risk models via
  machine-learning algorithms, Journal of Banking \& Finance 34~(11) (2010)
  2767--2787.

\bibitem{raissi2019physics}
M.~Raissi, P.~Perdikaris, G.~E. Karniadakis, Physics-informed neural networks:
  A deep learning framework for solving forward and inverse problems involving
  nonlinear partial differential equations, Journal of Computational Physics
  378 (2019) 686--707.

\bibitem{mao2020physics}
Z.~Mao, A.~D. Jagtap, G.~E. Karniadakis, Physics-informed neural networks for
  high-speed flows, Computer Methods in Applied Mechanics and Engineering 360
  (2020) 112789.

\bibitem{shukla2020physics}
K.~Shukla, P.~C. Di~Leoni, J.~Blackshire, D.~Sparkman, G.~E. Karniadakis,
  Physics-informed neural network for ultrasound nondestructive quantification
  of surface breaking cracks, Journal of Nondestructive Evaluation 39~(3)
  (2020) 1--20.

\bibitem{sahli2020physics}
F.~Sahli~Costabal, Y.~Yang, P.~Perdikaris, D.~E. Hurtado, E.~Kuhl,
  Physics-informed neural networks for cardiac activation mapping, Frontiers in
  Physics 8 (2020) 42.

\bibitem{yin2021non}
M.~Yin, X.~Zheng, J.~D. Humphrey, G.~E. Karniadakis, Non-invasive inference of
  thrombus material properties with physics-informed neural networks, Computer
  Methods in Applied Mechanics and Engineering 375 (2021) 113603.

\bibitem{waheed2020eikonal}
U.~Waheed, E.~Haghighat, T.~Alkhalifah, C.~Song, Q.~Hao, Eikonal solution using
  physics-informed neural networks, in: 82nd EAGE Annual Conference \&
  Exhibition, Vol. 2020, European Association of Geoscientists \& Engineers,
  2020, pp. 1--5.

\bibitem{shukla2021physics}
K.~Shukla, A.~D. Jagtap, J.~L. Blackshire, D.~Sparkman, G.~E. Karniadakis, A
  physics-informed neural network for quantifying the microstructure properties
  of polycrystalline nickel using ultrasound data, arXiv preprint
  arXiv:2103.14104 (2021).

\bibitem{VPINN}
E.~Kharazmi, Z.~Zhang, G.~E. Karniadakis, Variational physics-informed neural
  networks for solving partial differential equations, arXiv preprint
  arXiv:1912.00873.

\bibitem{cai2021flow}
S.~Cai, Z.~Wang, F.~Fuest, Y.~J. Jeon, C.~Gray, G.~E. Karniadakis, Flow over an
  espresso cup: inferring 3-d velocity and pressure fields from tomographic
  background oriented schlieren via physics-informed neural networks, Journal
  of Fluid Mechanics 915.

\bibitem{cai2021artificial}
S.~Cai, H.~Li, F.~Zheng, F.~Kong, M.~Dao, G.~E. Karniadakis, S.~Suresh,
  Artificial intelligence velocimetry and microaneurysm-on-a-chip for
  three-dimensional analysis of blood flow in physiology and disease,
  Proceedings of the National Academy of Sciences 118~(13).

\bibitem{jagtap2020conservative}
A.~D. Jagtap, E.~Kharazmi, G.~E. Karniadakis, Conservative physics-informed
  neural networks on discrete domains for conservation laws: Applications to
  forward and inverse problems, Computer Methods in Applied Mechanics and
  Engineering 365 (2020) 113028.

\bibitem{jagtap2020extended}
A.~D. Jagtap, G.~E. Karniadakis, Extended physics-informed neural networks
  (xpinns): A generalized space-time domain decomposition based deep learning
  framework for nonlinear partial differential equations, Communications in
  Computational Physics 28~(5) (2020) 2002--2041.

\bibitem{hpVPINN}
E.~Kharazmi, Z.~Zhang, G.~E. Karniadakis, hp-vpinns: Variational
  physics-informed neural networks with domain decomposition, Computer Methods
  in Applied Mechanics and Engineering 374 (2021) 113547.

\bibitem{sergeev2018horovod}
A.~Sergeev, M.~Del~Balso, Horovod: fast and easy distributed deep learning in
  tensorflow, arXiv preprint arXiv:1802.05799.

\bibitem{hennigh2020nvidia}
O.~Hennigh, S.~Narasimhan, M.~A. Nabian, A.~Subramaniam, K.~Tangsali,
  M.~Rietmann, J.~d.~A. Ferrandis, W.~Byeon, Z.~Fang, S.~Choudhry, Nvidia
  simnet\^{}$\{$TM$\}$: an ai-accelerated multi-physics simulation framework,
  arXiv preprint arXiv:2012.07938.

\bibitem{goyal2017accurate}
P.~Goyal, P.~Doll{\'a}r, R.~Girshick, P.~Noordhuis, L.~Wesolowski, A.~Kyrola,
  A.~Tulloch, Y.~Jia, K.~He, Accurate, large minibatch sgd: Training imagenet
  in 1 hour, arXiv preprint arXiv:1706.02677.

\bibitem{DeepSpeed}
{DeepSpeed}, \url{https://github.com/microsoft/DeepSpeed}, accessed:
  2020-06-12.

\bibitem{rasley2020deepspeed}
J.~Rasley, S.~Rajbhandari, O.~Ruwase, Y.~He, Deepspeed: System optimizations
  enable training deep learning models with over 100 billion parameters, in:
  Proceedings of the 26th ACM SIGKDD International Conference on Knowledge
  Discovery \& Data Mining, 2020, pp. 3505--3506.

\bibitem{xu2020distributed}
K.~Xu, W.~Zhu, E.~Darve, Distributed machine learning for computational
  engineering using mpi, arXiv preprint arXiv:2011.01349.

\bibitem{alnaes2015fenics}
M.~Aln{\ae}s, J.~Blechta, J.~Hake, A.~Johansson, B.~Kehlet, A.~Logg,
  C.~Richardson, J.~Ring, M.~E. Rognes, G.~N. Wells, The fenics project version
  1.5, Archive of Numerical Software 3~(100).

\bibitem{jagtap2020adaptive}
A.~D. Jagtap, K.~Kawaguchi, G.~E. Karniadakis, Adaptive activation functions
  accelerate convergence in deep and physics-informed neural networks, Journal
  of Computational Physics 404 (2020) 109136.

\bibitem{jagtap2020locally}
A.~D. Jagtap, K.~Kawaguchi, G.~Em~Karniadakis, Locally adaptive activation
  functions with slope recovery for deep and physics-informed neural networks,
  Proceedings of the Royal Society A 476~(2239) (2020) 20200334.

\bibitem{baydin2018automatic}
A.~G. Baydin, B.~A. Pearlmutter, A.~A. Radul, J.~M. Siskind, Automatic
  differentiation in machine learning: a survey, Journal of machine learning
  research 18.

\bibitem{tang2020review}
H.~Tang, R.~Haynes, G.~Houzeaux, A review of domain decomposition methods for
  simulation of fluid flows: Concepts, algorithms, and applications, Archives
  of Computational Methods in Engineering (2020) 1--33.

\bibitem{dolean2015introduction}
V.~Dolean, P.~Jolivet, F.~Nataf, An introduction to domain decomposition
  methods: algorithms, theory, and parallel implementation, SIAM, 2015.

\bibitem{gropp2014using}
W.~Gropp, T.~Hoefler, R.~Thakur, E.~Lusk, Using advanced MPI: Modern features
  of the message-passing interface, MIT Press, 2014.

\bibitem{lonvcar2016openmp}
V.~Lon{\v{c}}ar, L.~E. Young-S, S.~{\v{S}}krbi{\'c}, P.~Muruganandam, S.~K.
  Adhikari, A.~Bala{\v{z}}, Openmp, openmp/mpi, and cuda/mpi c programs for
  solving the time-dependent dipolar gross--pitaevskii equation, Computer
  physics communications 209 (2016) 190--196.

\bibitem{ruder2016overview}
S.~Ruder, An overview of gradient descent optimization algorithms, arXiv
  preprint arXiv:1609.04747.

\bibitem{le2011optimization}
Q.~V. Le, J.~Ngiam, A.~Coates, A.~Lahiri, B.~Prochnow, A.~Y. Ng, On
  optimization methods for deep learning, in: ICML, 2011.

\bibitem{dean2012large}
J.~Dean, G.~S. Corrado, R.~Monga, K.~Chen, M.~Devin, Q.~V. Le, M.~Z. Mao,
  M.~Ranzato, A.~Senior, P.~Tucker, et~al., Large scale distributed deep
  networks.

\bibitem{byrd1995limited}
R.~H. Byrd, P.~Lu, J.~Nocedal, C.~Zhu, A limited memory algorithm for bound
  constrained optimization, SIAM Journal on scientific computing 16~(5) (1995)
  1190--1208.

\bibitem{ghia}
U.~Ghia, K.~N. Ghia, C.~Shin, High-re solutions for incompressible flow using
  the navier-stokes equations and a multigrid method, Journal of computational
  physics 48~(3) (1982) 387--411.

\bibitem{daghaghi2021accelerating}
S.~Daghaghi, N.~Meisburger, M.~Zhao, A.~Shrivastava, Accelerating slide deep
  learning on modern cpus: Vectorization, quantizations, memory optimizations,
  and more, Proceedings of Machine Learning and Systems 3.

\bibitem{Rusu_ICRA2011_PCL}
R.~B. Rusu, S.~Cousins, {3D is here: Point Cloud Library (PCL)}, in: {IEEE
  International Conference on Robotics and Automation (ICRA)}, Shanghai, China,
  2011.

\bibitem{METIS}
G.~Karypis, V.~Kumar, {MeTis: Unstructured Graph Partitioning and Sparse Matrix
  Ordering System, Version 4.0}.

\end{thebibliography}
\end{document}